\documentclass[usenatbib]{mnras}

\usepackage{newtxtext,newtxmath}

\usepackage[T1]{fontenc}
\usepackage{ae,aecompl}

\usepackage{graphicx}	% Including figure files
\usepackage{amsmath}	% Advanced maths commands
\usepackage{amssymb}	% Extra maths symbols
\usepackage{relsize}
\usepackage{tikz}
\usetikzlibrary{shapes.geometric, arrows}
\usepackage{booktabs}
\usepackage{caption}
\usepackage{subcaption}
\usepackage{float}
\usepackage{pdflscape}
\usepackage{afterpage}
\usepackage{mathtools}

 \newcommand{\kepler}{\textit{Kepler}}
 \newcommand{\corot}{\textit{CoRoT}}
 \newcommand{\ktwo}{\textit{K2}}
 \newcommand{\gaia}{Gaia}
 \newcommand{\ngts}{NGTS}
 \newcommand{\mearth}{MEarth}
 \newcommand{\kreutzer}{Kreutzer et al. [submitted]}
 \newcommand{\gacf}{G-ACF}
 \newcommand{\casutools}{\textsc{casutools}}
 \newcommand{\teff}{$T_{\text{eff}}$}
 \newcommand{\bprp}{$G_{BP} - G_{RP}$}
 \newcommand{\grp}{$G - G_{RP}$}
 \newcommand{\nobjs}{$829,481$}
 \newcommand{\nperiods}{$16,880$}

\tikzstyle{outputbad} = [rectangle, rounded corners, minimum width=3cm, minimum height=1cm,text centered, draw=black, fill=red!30]
\tikzstyle{outputgood} = [rectangle, rounded corners, minimum width=3cm, minimum height=1cm,text centered, draw=black, fill=green!30]
\tikzstyle{io} = [trapezium, trapezium left angle=70, trapezium right angle=110, minimum width=3cm, minimum height=1cm, text centered, draw=black, fill=blue!30]
\tikzstyle{process} = [rectangle, minimum width=3cm, minimum height=1cm, text centered, draw=black, fill=orange!30]
\tikzstyle{decision} = [diamond, minimum width=3cm, minimum height=1cm, text centered, draw=black, fill=purple!30]
\tikzstyle{arrow} = [thick,->,>=stealth]

\title[\ngts{} Variability Study]{Periodic stellar variability from almost a million \ngts{} light curves.}

\author[J. T. Briegal et al.]{
\parbox{\textwidth}{Joshua~T.~Briegal,$^{1}$\thanks{E-mail: jtb34@cam.ac.uk}
Edward Gillen,$^{2,1}$\thanks{Winton Fellow}
Didier Queloz,$^{1}$
Simon Hodgkin,$^{3}$
Jack S.~Acton,$^{4}$
David~R.~Anderson,$^{5,6}$
David~J.~Armstrong,$^{5,6}$
Matthew~P.~Battley,$^{5,6}$
Daniel Bayliss,$^{5,6}$
Matthew~R.~Burleigh,$^{4}$
Edward~M.~Bryant,$^{5,6}$
Sarah~L.~Casewell,$^{4}$
Jean~C.~Costes,$^{7}$
Philipp~Eigm\"uller,$^{8}$,
Samuel Gill,$^{5,6}$
Michael~R.~Goad,$^{4}$
Maximilian~N.~G{\"u}nther,$^{9,14}$\thanks{Juan Carlos Torres Fellow}\thanks{ESA Research Fellow}
Beth~A.~Henderson,$^{4}$
James A. G. Jackman,$^{10,5,6}$
James~S.~Jenkins,$^{11,12}$
Lars T. Kreutzer, $^{1,15,16}$
Maximiliano Moyano, $^{17}$
Monika~Lendl,$^{13}$
Gareth D. Smith,$^{1}$
Rosanna H.~Tilbrook,$^{4}$
Christopher A.~Watson,${^7}$
Richard~G.~West,$^{5,6}$
Peter~J.~Wheatley$^{5,6}$
}
\\
\\
\parbox{\textwidth}{
$^{1}$Astrophysics Group, Cavendish Laboratory, J.J. Thomson Avenue, Cambridge CB3 0HE, UK.\\
$^{2}$Astronomy Unit, Queen Mary University of London, Mile End Road, London E1 4NS, UK\\
$^{3}$Institute of Astronomy, University of Cambridge, Madingley Rise, Cambridge CB3 0HA, UK.\\
$^{4}$School of Physics and Astronomy, University of Leicester, University Road, Leicester, LE1 7RH, UK\\
$^{5}$Department of Physics, University of Warwick, Gibbet Hill Road, Coventry CV4 7AL, UK\\
$^{6}$Centre for Exoplanets and Habitability, University of Warwick, Gibbet Hill Road, Coventry CV4 7AL, UK\\
$^{7}$Astrophysics Research Centre, School of Mathematics and Physics, Queen's University Belfast, BT7 1NN Belfast, UK\\
$^{8}$Institute of Planetary Research, German Aerospace Center, Rutherfordstrasse 2., 12489 Berlin, Germany\\
$^{9}$Department of Physics and Kavli Institute for Astrophysics and Space Research, Massachusetts Institute of Technology, Cambridge, MA 02139, USA\\
$^{10}$School of Earth and Space Exploration, Arizona State University, Tempe, AZ 85287, USA\\
$^{11}$N\'ucleo de Astronom\'ia, Facultad de Ingenier\'ia y Ciencias, Universidad Diego Portales, Av. Ej\'ercito 441, Santiago, Chile\\
$^{12}$Centro de Astrof\'isica y Tecnolog\'ias Afines (CATA), Casilla 36-D, Santiago, Chile\\
$^{13}$Observatoire de Gen{\`e}ve, Universit{\'e} de Gen{\`e}ve, 51 Chemin Pegasi, 1290 Sauverny, Switzerland\\
$^{14}$ European Space Agency (ESA), European Space Research and Technology Centre (ESTEC), Keplerlaan 1, 2201 AZ Noordwijk, The Netherlands \\
$^{15}$Department of Applied Mathematics and Theoretical Physics, University of Cambridge, Cambridge, CB3 OWA, UK\\
$^{16}$Max Planck Institute for Gravitational Physics (Albert Einstein Institute), Am M\"uhlenberg 1, 14476 Golm, Germany\\
$^{17}$Instituto de Astronom\'ia, Universidad Cat\'olica del Norte,Angamos 0610, 1270709, Antofagasta, Chile\\
}  
}

\date{Accepted XXX. Received YYY; in original form ZZZ}

\pubyear{2021}

\begin{document}
\label{firstpage}
\pagerange{\pageref{firstpage}--\pageref{lastpage}}
\maketitle

\begin{abstract}
We analyse \nobjs{} stars from the Next Generation Transit Survey (\ngts{}) to extract variability periods. We utilise a generalisation of the autocorrelation function (the \gacf{}), which applies to irregularly sampled time series data. We extract variability periods for \nperiods{} stars from late-A through to mid-M spectral types {and periods between $\sim$\,0.1 and 130 days} with no assumed variability model. 
We find variable signals associated with a number of astrophysical phenomena, including stellar rotation, pulsations and multiple-star systems.
The extracted variability periods are compared with stellar parameters taken from \gaia{} DR2, which allows us to identify distinct regions of variability in the Hertzsprung-Russell Diagram. We explore a sample of {rotational} main-sequence objects in period-colour space, in which we observe a dearth of rotation periods between 15 and 25 days. This `bi-modality' was previously only seen in space-based data.
We demonstrate that stars in sub-samples above and below the period gap appear to arise from a stellar population not significantly contaminated by excess multiple systems.
We also observe a small population of long-period variable M-dwarfs, which highlight a departure from the predictions made by rotational evolution models fitted to solar-type main-sequence objects. 
The \ngts{} data spans a period and spectral type range that links previous rotation studies such as those using data from \kepler{}, \ktwo{} and \mearth{}.

\end{abstract}


%%%%%%%%%%%%%%%%%%%%%%%%%%%%%%%%%%%%%%%%%%%%%%%%%%

%%%%%%%%%%%%%%%%% BODY OF PAPER %%%%%%%%%%%%%%%%%%

\section{Introduction}

{Many of a star's physical properties can be inferred from its brightness variations over time.} This variability can arise from a number of mechanisms, {either} intrinsic to the star through changing physical properties of the star and its photosphere, {or} through external factors such as orbiting bodies and discs. The rotation of magnetically active stars will also cause visible brightness changes.
\citet{Eyer2008} categorises a large number of distinct variability types which range in period from milliseconds to centuries and in amplitude from a few parts-per-million (ppm) to orders of magnitude in the most explosive forms of variability.

Stellar rotation can be measured through photometric observation, as magnetic surface activity such as spots and plages cause photometric brightness fluctuations over time that is modulated by both the rotation of active regions across the star, as well as active region evolution. Constraining stellar rotation rates is important, as this provides insight into the angular momentum of the star. \citet{Skumanich1972} first hypothesised that a star's rotation rate could be age dependant, obtaining the empirical relation between rotation period $P_{\mathrm{rot}}$ and age $t$: $P_{\mathrm{rot}} \propto t^{0.5}$. Knowing a star's age is fundamental to fully understanding its evolutionary state, and so being able to infer this property from an observable quantity such as rotation would greatly improve our understanding of stars in the local neighbourhood.
In \citet{Barnes2003} a semi-empirical model for deriving stellar ages from colour and rotation period was suggested, and the term `gyrochronology' was coined. This model was subject to further improvements in \citet{Barnes2007}, a model which is commonly still used to age Solar-type and late-type main-sequence stars. These models work especially well for stars older than the age of the Hyades cluster, by which time we expect the initial angular momentum of stars to have little effect on the rotation period, and the angular momentum evolution to follow a Skumanich law \citep{Kawaler1988}. For low mass stars, it is widely accepted that late-time angular momentum loss will be governed by magnetised stellar winds which depend on magnetic field topology and stellar mass \citep{Booth2017}.
For young stars ($< 10$ Myr) angular momentum evolution may be dependant on magnetic coupling between the star and disc. Studies of pre-main-sequence stars in young clusters such as T-Tauri stars in the Taurus-Auriga molecular cloud \citep{Hartmann1989} or in NGC 2264 \citep{Sousa2016} show high levels of short period ($< 10$ day) photometric variability, {but objects with circumstellar discs present appear to rotate slower than those without, highlighting the effect of star-disc coupling on angular momentum evolution.}

Understanding a star's activity is important for exoplanet surveys. Not only is stellar activity a large source of noise in both transit and RV surveys \citep[e.g.][]{Queloz2001, Haywood2014, Dumusque2017}, but stellar activity may also influence the potential habitability of orbiting planets.
Stars that rotate rapidly, for example, often display higher flare rates than their more slowly rotating cousins, and these flares can be important for potential exoplanet habitability. On the one hand, flares can erode exoplanet atmospheres and modify their chemistry \citep[e.g.][]{Segura2010,Seager2013,Tilley2019}, while on the other, they can help initiate prebiotic chemistry and seed the building blocks of life \citep{Ranjan2017,Rimmer2018}, which may be especially important for M dwarf systems.

The angular momentum of a host star and its planets are intrinsically linked. \citet{Gallet2018} demonstrate that tidal interactions between a host star and a close-in planet can affect the surface rotation of the star. They observe a deviation in rotation period from the expected magnetic braking law during the early MS phase of low-mass stars in the Pleiades cluster, which {the authors attribute} to planetary engulfment events. Conversely, angular momentum transfer through tidal interactions must be considered in the context of stellar spin-down through magnetic braking. The analysis by \citet{Damiani2015} demonstrates that to accurately model tidal dissipation efficiency and orbital migration the stellar angular momentum loss through magnetised stellar winds must be accounted for.

% Extremely active stars may not allow atmospheres suitable for life to form due to strong magnetic stripping of secondary atmospheres around rocky planets. \citep{Kay2016}.

Large scale photometric variability studies have recently allowed for data-driven analysis of stellar variability in extremely large samples. Stellar clusters allow studies of groups of stars with similar formation epochs and evolutionary conditions, so historically have been targeted by systematic surveys. These observations have come from ground-based surveys such as Monitor \citep{Hodgkin2006, Aigrain2007} with observations of NGC 2516 \citep{Irwin2007}, SuperWASP \citep{Pollacco2006} with observations of the Coma Berenices open cluster \citep{CollierCameron2009} and HATNet \citep{Bakos2004} with observations of FGK Pleiades stars \citep{Hartman2010}. Recently, \ngts{} \citep{Wheatley2018} observed the $\sim 115$ Myr old cluster Blanco 1, and a study by \citet{Gillen2020} demonstrated a well-defined single-star rotation sequence {which was also observed by KELT \citep{Pepper2012} and studied in \citet{Cargile2014}. In both of these works a} similar sequence was observed for stars in the similarly aged Pleiades, indicating angular momentum evolution of mid-F to mid-K stars follows a well-defined pathway which is strongly imprinted by $\sim 100$ Myr. 

{As part of the transient search conducted by the All-Sky Automated Survey for Supernovae \citep[ASAS-SN;][]{Shappee2014}, a catalogue of observed variable stars has been compiled. This catalogue contains variability periods and classifications for 687,695 objects\footnote{Accessed on 09/11/2021} taken from a series of publications entitled `The ASAS-SN catalogue of variable stars' \citep[e.g.][]{Jayasinghe2018, Jayasinghe2021}. Such catalogues are not focused on specific clusters or stellar types, but provide a broad view of different forms of stellar variability.}

Space missions have allowed wide-field photometric variability surveys of stars with high precision and {continuous} time coverage. \corot{} \citep{Auvergne2009}, \kepler{} \citep{Borucki2010}, the extended \kepler{} mission \citep[\ktwo{};][]{Howell2014} {and TESS \citep{Ricker2014}} have provided a wealth of stellar photometric data, which in turn has been the subject of extensive rotation studies \citep{Ciardi2011, Basri2011, Affer2012, McQuillan2014, Davenport2018, CantoMartins2020, Gordon2021}, revealing large scale trends in stellar variability periods. In particular, studies by \citet{McQuillan2014} and \citet{Davenport2018} demonstrated a distinct bi-modal structure in the rotation periods of main-sequence stars with respect to colour. \citet{Gordon2021} followed up these studies with analysis of data from the \ktwo{} mission, hypothesising the bi-modal structure arises from a broken spin-down law, caused by an internal angular momentum transfer between the core and convective envelope. Further details of this model are discussed in Section \ref{sec:discussion}.

The Next Generation Transit Survey \citep[\ngts{};][]{Wheatley2018} is a  ground-based wide-field photometric survey achieving routinely milli-magnitude range photometric precision  with 12-second sampling cadence and long observation baselines, typically 250 nights of data per target field. The primary science goal of \ngts{} is to extend the wide-field ground-based detection of transiting exoplanets to at least the Neptune size range. Such high precision photometry lends itself well to ancillary stellar physics such as cluster rotation analysis \citep{Gillen2020} or stellar-flare detection and characterisation \citep{Jackman2019}.
Ground-based observation adds extra layers of difficultly in variability studies when compared to space telescope data, as we must consider irregular sampling and telluric effects. In this study, we employ a generalisation of the autocorrelation function (the \gacf{}) which applies to this irregular sampling. We elect to use an autocorrelation function to extract variability as this has proven to be successful for extracting stellar variability by \citet{McQuillan2013, McQuillan2014} \& \citet{Angus2018} and for \ngts{} data in \citet{Gillen2020}. An Autocorrelation Function (ACF) also allows better detection of pseudo-periodic and phase-shifting variability often seen in young, active stars in comparison to more rigid variability extraction techniques {such as Lomb-Scargle periodograms} \citep{Lomb1976, Scargle1982}.

The paper is organised as follows.
In section \ref{sec:ngts} we discuss the Next Generation Transit Survey and the data used, and in section \ref{sec:methods} we outline the methods used in this study to extract rotation periods. Our results are summarised in section \ref{sec:results}, with a discussion of these results in section \ref{sec:discussion} and a brief summary of our findings in section \ref{sec:conclusion}.

\section{The Next Generation Transit Survey (\ngts{})}
\label{sec:ngts}

\ngts{} is an array of twelve 20cm telescopes based at ESO's Paranal Observatory in Chile. Each telescope is coupled to a 2K $\times$ 2K e2V deep-depleted Andor IKon-L CCD camera with 13.5 $\mu$m pixels, corresponding to an on-sky size of $4.97"$. The data for this study were taken with the array in \emph{survey mode}, where each telescope observes a sequence of survey fields (generally 2 per night), each field having an on-sky size of $\sim 8$ deg$^2$. These fields are spaced such as one field rises above 30\textdegree\ the previous field sets below 30\textdegree. This typically results in approximately 500 hours coverage per field spread over 250 nights.

Fields are selected based on the density of stars, the proportion of dwarf stars, the ecliptic latitude and proximity to any bright or extended objects. Fields are typically selected with $\leq$ 15,000 stars brighter than an $I$ band magnitude of 16, of which $\geq$ 70\% are dwarf stars. These fields will be more than 20\textdegree\ from the Galactic plane. Fields within 30\textdegree\ of the ecliptic plane are also avoided due to the Moon affecting readings {during about three nights per month}.

In this study, we use the {final light curves, associated metadata and quality flags} of the standard \ngts{} pipeline as described in \citet{Wheatley2018}. The data for each field is reduced and photometric measurements are made on source apertures to assemble a light curve per target star. As a part of the pipeline, these light curves are passed through a custom implementation of the SysRem algorithm \citep{Tamuz2005} which removes signals common to multiple stars arising from various sources including the instrument, reduction software and the atmosphere.

\subsection{Light Curve Extraction}
\label{subsec:lc_extraction}

Photometric light curves are extracted for all sources detected within each \ngts{} field. Source detection is done using the \textsc{imcore} module in \casutools{} \citep{Irwin2004} to generate an object list that is cross-matched against a number of catalogues. 
\ngts{} generates its own input source catalogue, as explained in Section 5 of \citet{Wheatley2018}. This source catalogue is cross-matched against a number of external catalogues. Cross-matching is done in position, as well as in colour and separation to limit spurious matches. This allows flagging of potential unresolved binaries in \ngts{} apertures. 

A soft-edged circular aperture with a radius of 3 pixels (15 arcsec) is placed over each of these sources and placed in pixel coordinates using per-image astrometric solutions to account for radial distortion in addition to the autoguiding system. The sky background is estimated using bilinear interpolation of a grid of $64 \times 64$ pixel regions for which the sky level is determined using a k-sigma clipped median.
These raw light curves are then passed through the detrending pipeline described in section \ref{subsec:detrending} (Section 6 of \citet{Wheatley2018}).

\subsection{Systematics Correction}
\label{subsec:detrending}

In order to correct first-order offsets common to all light curves the detrending algorithm calculates a mean light curve for all objects to be used as an artificial `standard star'. 
This detrending algorithm is based on the SysRem algorithm first described in \citet{Tamuz2005} and adapted from the WASP project \citep{CollierCameron2006}.
Of note, this approach may not fully remove systematic trends which correlate with Moon phase and sidereal time. In particular, Moon phased signals will show artefacts of imperfect background subtraction and any non-linearity in the detectors.  

\subsection{Data Selection}
\label{subsec:data_selection}

The \ngts{} pipeline provides flags per image and per timestamp per object light curve which we use to pre-process light curves for variability analysis. These flags alert us to bad-quality data points as a result of pixel saturation, blooming spikes from nearby bright sources, cosmics and other crossing events (including weather and laser guide stars) and any sky background changes.
We removed any flagged data point from our light curves, and additionally checked if the majority of the light curve had been flagged ($>80\%$ of data points). If this was the case, we removed the objects from processing.

We clipped our flux data to remove any points lying further than 3 median-absolute-deviations (MAD) from the median to remove any outliers not caught by the \ngts{} pipeline flags. This cut may remove some variability signals such as long-period eclipsing binaries where the variability is a small fraction of the phase curve. Manual inspection of a single field confirmed this was not the case, however, this cannot be guaranteed for all fields processed.
Finally, in order to speed up data processing, we binned our light curve into 20-minute time bins. This reduced the number of data points to process per light curve from 200,000 to roughly 10,000. The \gacf{} {computation time} scales as $\mathcal{O}(n^2m)$ for $n$ data points with $m$ lag time steps, so reducing the number of timestamps in our light curve {significantly} improved processing time with a caveat that we will be unable to detect any periods below 40 minutes. For this study that is focused on longer period variability, this limit is not of concern.

\begin{figure}
    \centering
    \includegraphics[width=\linewidth]{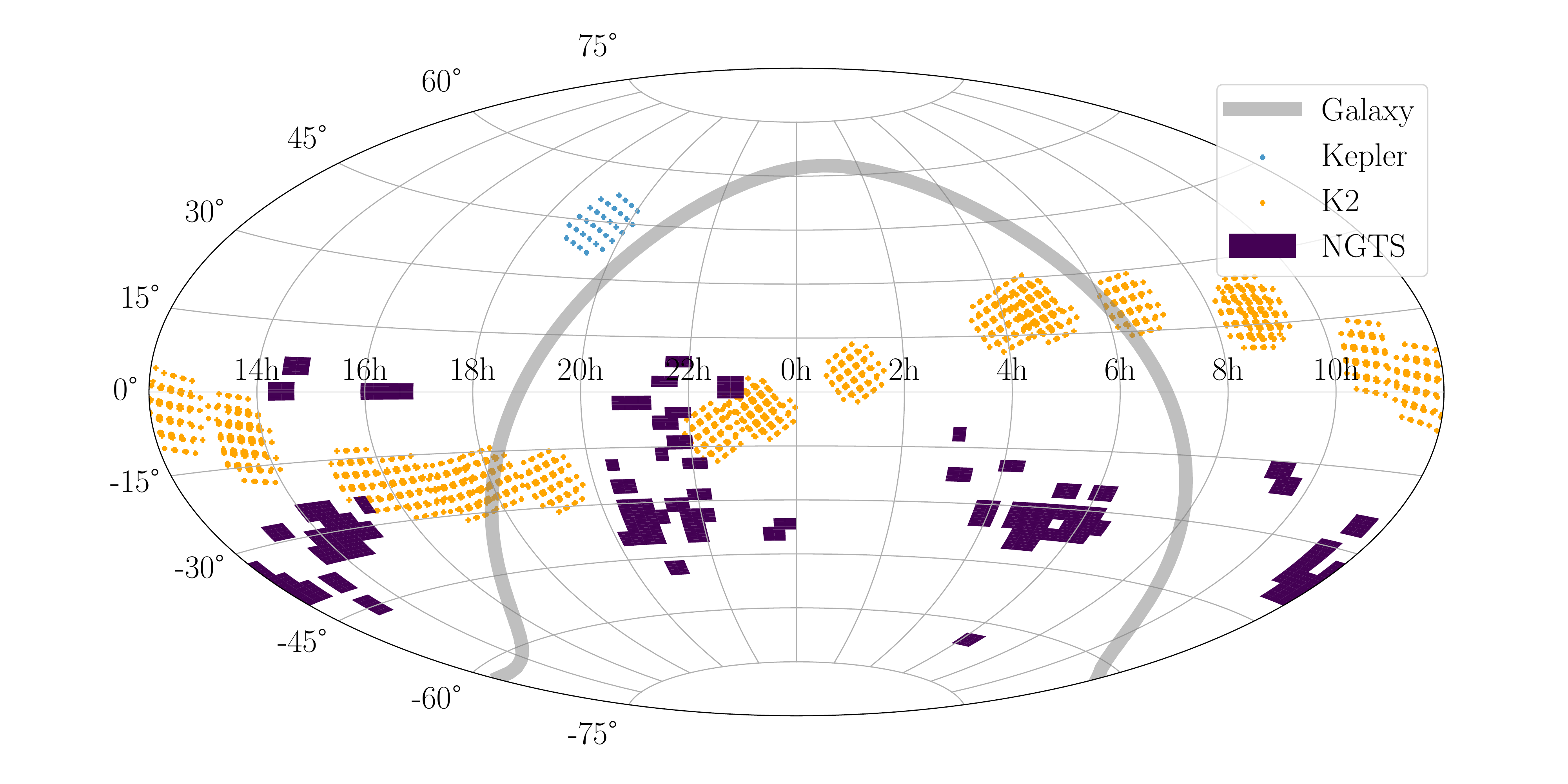}
    \caption{An ICRS plot of the position of the 94 NGTS fields used in this study (solid dark blue squares). The \kepler{} and \ktwo{} fields are included as blue and orange squares, respectively, as well as the galactic plane as a thick grey line.}
    \label{fig:field_positions}
\end{figure}

We removed 6 fields identified as containing large open cluster populations. This study will focus on stars in the field and this avoids contamination of large numbers of young variable stars in known {open clusters}. Removing these 6 fields left a total of \nobjs{} light curves to process. The positions of the 94 NGTS fields in RA and Dec used in this study are shown in Figure \ref{fig:field_positions}. In this Figure, we plot the \kepler{} and \ktwo{} field centre pointings, as well as the position of the galactic plane.

{The 94 fields used in this study were observed for an average of 141 nights during different observation campaigns (lasting an average of 218 days) between September 2015 and November 2018. The shortest observational baseline for this data set was 84 days and the longest 272 days. 73 of the 94 fields had observational baselines over 200 days.}
{We detected periodic variability in light curves spanning $8 < I_{\text{NGTS}} < 16$ mag with 50\% (90\%) of our detections being brighter than 13.5 (15.4) mag.}

\section{Period Detection}
\label{sec:methods}

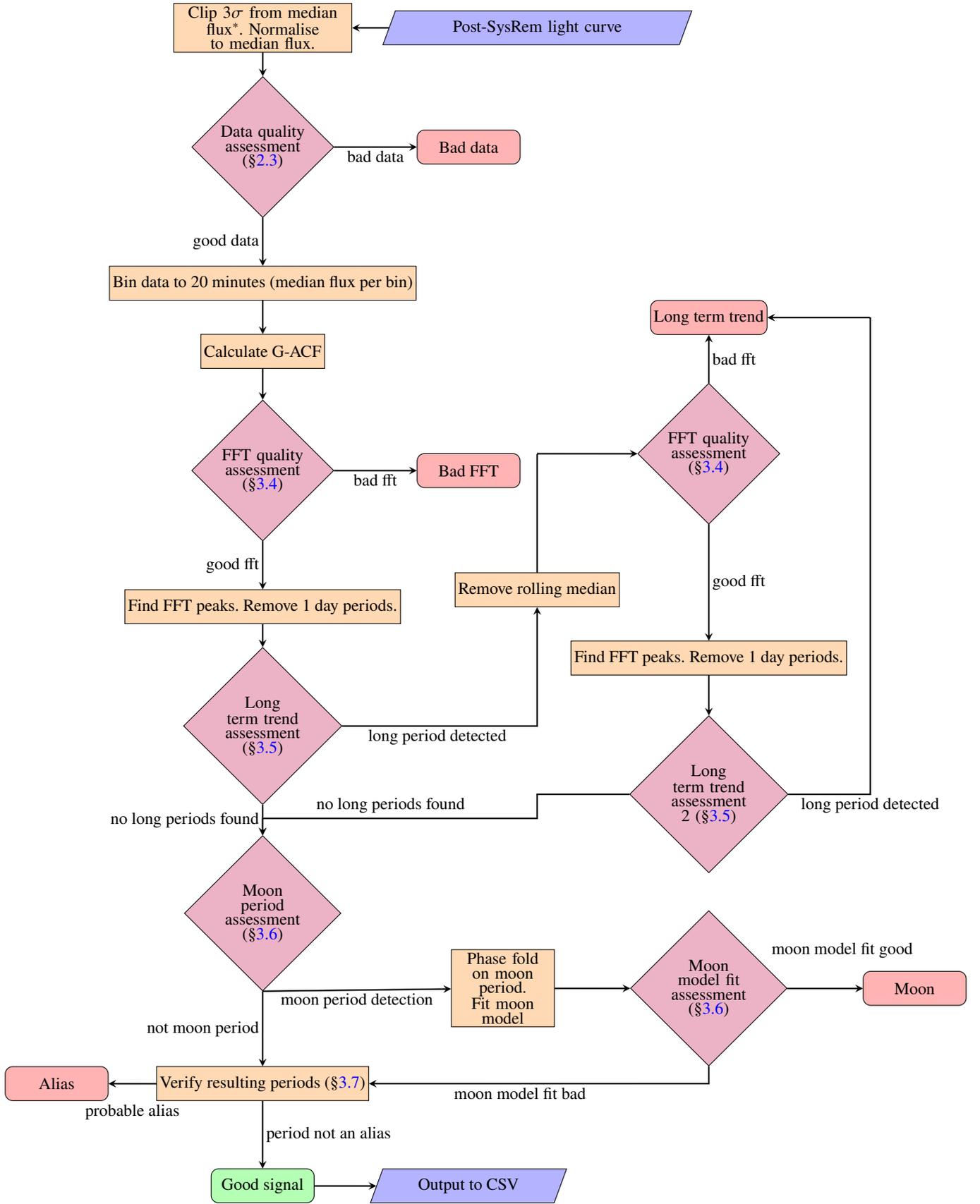
\begin{figure*}
\tikzset{
  font={\fontsize{14pt}{12}\selectfont}}
\begin{tikzpicture}[node distance=2cm]
\begin{scope}[scale=0.64, transform shape]
\node (lc) [io, text width=5cm] {Post-SysRem light curve};
\node (sigma) [process, left of=lc, text width=5cm, xshift=-6cm] {Clip 3$\sigma$ from median flux$^{*}$. Normalise to median flux.};
\node (dataqual) [decision, below of=sigma, yshift=-1.5cm, text width=2.5cm] {Data quality assessment (\S\ref{subsec:data_selection})};

\node (baddata) [outputbad, right of=dataqual, xshift=4cm] {Bad data};

\node (bin) [process, below of=dataqual, yshift=-2cm] {Bin data to 20 minutes (median flux per bin)};
\node (gacf) [process, below of=bin] {Calculate \gacf};
\node (checkfft) [decision, below of=gacf, yshift=-1.5cm, text width=2.5cm] {FFT quality assessment (\S\ref{subsec:fft})};

\node (badfft) [outputbad, right of=checkfft, xshift=4cm] {Bad FFT};

\node (peaks) [process, below of=checkfft, yshift=-2cm] {Find FFT peaks. Remove 1 day periods.};
\node (longterm) [decision, below of=peaks, yshift=-1.5cm, text width=2.5cm] {Long term trend assessment (\S\ref{subsec:longtermtrend})};

\node (removemedian) [process, right of=longterm, xshift=6cm, yshift=4cm] {Remove rolling median};
\node (fftcheckltt) [decision, right of=removemedian, xshift=3cm, yshift=4cm, text width=2.5cm] {FFT quality assessment (\S\ref{subsec:fft})};
\node (badltt) [outputbad, above of=fftcheckltt, yshift=2cm] {Long term trend};
\node (peaksltt) [process, below of=fftcheckltt, yshift=-4cm] {Find FFT peaks. Remove 1 day periods.};
\node (longtermltt) [decision, below of=peaksltt, yshift=-2cm, text width=2.5cm] {Long term trend assessment 2 (\S\ref{subsec:longtermtrend})};

\node (moon) [decision, below of=longterm, yshift=-3.5cm, text width=2.5cm] {Moon period assessment (\S\ref{subsec:moon})};
\node (moonmodel) [process, right of=moon, yshift=-2.2cm, xshift=5cm, text width=2.5cm] {Phase fold on moon period. Fit moon model};
\node (moonmodelok) [decision, right of=moonmodel, xshift=4cm, text width=2.5cm] {Moon model fit assessment (\S\ref{subsec:moon})};
\node (moonbad) [outputbad, right of=moonmodelok, xshift=4cm] {Moon};

\node (verify) [process, below of=moon, yshift=-3cm] {Verify resulting periods (\S\ref{subsec:aliases})};
\node (alias) [outputbad, left of=verify, xshift=-4cm] {Alias};
\node (ok) [outputgood, below of=verify, yshift=-1cm] {Good signal};
\node (output) [io, right of=ok, xshift=4cm] {Output to CSV};

\draw [arrow] (lc) -- (sigma);
\draw [arrow] (sigma) -- (dataqual);
\draw [arrow] (dataqual) -- node[anchor=north] {bad data} (baddata);
\draw [arrow] (dataqual) -- node[anchor=east] {good data} (bin);
\draw [arrow] (bin) -- (gacf);
\draw [arrow] (gacf) -- (checkfft);
\draw [arrow] (checkfft) -- node[anchor=north] {bad fft} (badfft);
\draw [arrow] (checkfft) -- node[anchor=east] {good fft} (peaks);
\draw [arrow] (peaks) -- (longterm);
\draw [arrow] (longterm) -| node[anchor=north, align=left, pos=0.2, xshift=0.5cm] {long period detected} ([yshift=-3cm, xshift=0cm]removemedian.south) -- (removemedian);
\draw [arrow] (longterm) -- node[anchor=east] {no long periods found} (moon);
\draw [arrow] (removemedian) |- ([xshift=0cm, yshift=4cm]removemedian.center) -- (fftcheckltt);
\draw [arrow] (fftcheckltt) -- node[anchor=west] {bad fft} (badltt);
\draw [arrow] (fftcheckltt) -- node[anchor=west] {good fft} (peaksltt);
\draw [arrow] (peaksltt) -- (longtermltt);
\draw [arrow] (longtermltt) -| node[anchor=north] {long period detected} ([xshift=3cm, yshift=0cm]badltt.east) |- (badltt);
\draw [arrow] (longtermltt.west) -| node[anchor=north east, xshift=-2cm] {no long periods found} ([yshift=0.5cm, xshift=8cm]moon.north) |- ([yshift=0.5cm, xshift=0cm]moon.north) -- (moon);
\draw [arrow] (moon.south) -- node[anchor=east, pos=0.5] {not moon period} (verify);
\draw [arrow] (moon.south) -- node[anchor=north] {moon period detection} (moonmodel);
\draw [arrow] (moonmodel) -- (moonmodelok);
\draw [arrow] (moonmodelok) -- node[anchor=south, yshift=0.8cm, xshift=0.5cm] {moon model fit good} (moonbad);
\draw [arrow] (moonmodelok) |- ([xshift=15cm, yshift=0cm]verify.west) -- node[anchor=north] {moon model fit bad} (verify);
\draw [arrow] (verify) -- node[anchor=north, yshift=-0.5cm] {probable alias} (alias);
\draw [arrow] (verify) -- node[anchor=west] {period not an alias} (ok);
\draw [arrow] (ok) -- (output);
\end{scope}
\end{tikzpicture}
\caption{A schematic of the period detection pipeline, per \ngts{}{} light curve.
{$^{*}\sigma$ refers to the median absolute deviation (MAD).}}
\label{fig:flow_chart}
\end{figure*}

The period detection pipeline is outlined in the flowchart in Figure \ref{fig:flow_chart}. Further details of each step are given in the subsequent sections.

The open-source code of the periodicity detection pipeline can be found on GitHub\footnote{\url{https://github.com/joshbriegal/periodicity_detection}}.

\subsection{Generalised Autocorrelation Function (\gacf{})}
\label{sec:GACF}
The \gacf{} is essentially an extended and generalised form of the standard auto-correlation function (ACF) which can be applied to any time series, regardless of sampling. A complete and detailed mathematical description of this algorithm is available in {a separate paper by \kreutzer{}}. This generalisation is done by (a) generalising the lag time $k$ to a \emph{generalised lag} $\hat{k}$ which is a continuous variable within the range of our time series and (b) defining a \emph{selection function} $\hat{S}$ and a \emph{weight function} $\hat{W}$.

Taking a standard definition of the ACF \citep[e.g.][]{Shumway2006}: 

\begin{align}
\label{eqn:ACF}
  \rho &(k) := \frac{1}{N} \sum\limits_{i=1}^{i_{\text{max}}-k} (X_i - \langle X_I\rangle)\times(X_{i+k} - \langle X_I\rangle),
\end{align}
where $\langle X_I\rangle$ denotes the mean of the time series values and the normalisation $N$ is the total sum of squares $N := \sum\limits_{i\in I} \left(X_i - \langle X_I\rangle\right)^2$. We can generalise this to:

\begin{align}
\label{eqn:\gacf{}}
  \hat{\rho}&\left(\hat{k}; \hat{W},\hat{S}\right) := \frac{1}{N}{\mathlarger{\mathlarger{\sum}}_{\mathclap{\substack{i\in I\\ t_i+\hat{k} \leq \max(T_I) }}}}\Bigl[  \Bigl(X(t_i) - \langle X_I\rangle\Bigr) \times \left(X(\hat{S}(t_i+\hat{k})) - \langle X_I\rangle\right)
  \nonumber \\
 & \qquad\qquad\qquad\quad\times\,\hat{W}\left(|\hat{S}(t_i + \hat{k})- (t_i + \hat{k}) |\right)\Bigr],
\end{align}
where $N := \sum\limits_{i\in I} \left(X_i - \langle X_I\rangle\right)^2$ denotes the total sum of squares and $\langle X_I\rangle$ is the mean of the time series values set.

\subsection{The selection function, $\hat{S}$}
\label{subsec:selection}

The selection function $\hat{S}$ provides a mapping between time labels within the original time series and the lagged time series at each lag time step. The most natural choice of a selection function would be to select the point closest in time within the original time series for each point in the lagged time series.
Further details of this selection function, including a cartoon outlining the method, are detailed in \kreutzer{}.

\subsection{The weight function, $\hat{W}$}
\label{subsec:weight}

The weight function, $\hat{W}$,  should be a function ${\hat{W} : [0,\infty) \to [0,1]}$ with $\hat{W}(0)\equiv 1$ to ensure that for a regular time series the \gacf{} reduces back to the standard ACF. One such example is a rational weight function such as
\begin{equation}
\label{eqn:fractional_weight_function}
 \hat{W}(\delta t) = \frac{1}{1+ \alpha \delta t} , \quad \alpha > 0,\, \delta t \geq 0,
\end{equation}
in which $\delta t$ is the time difference between the time label in the original time series and the lagged time series mapped by the selection function $\hat{S}$. The parameter $\alpha$ was taken as the median value of the time series (as a time difference from the first data point), as prescribed in \kreutzer{}. We experimented with two different weight functions and elected to use the rational function as the final extracted periods were not dependant on this choice and this function is very simple. We used the minimum gap between time stamps as our lag resolution (time steps in generalised lag, $\hat{k}$); this was 20 minutes as we bin the data prior to analysis (Section \ref{subsec:data_selection}).

% \begin{equation}
% \label{eqn:gaussian}
%  \hat{W}(\delta t) = -\frac{\delta t^2}{2\alpha^2} , \quad \alpha > 0,\, \delta t \geq 0
% \end{equation}

\subsection{Fast Fourier Transform (FFT)}
\label{subsec:fft}

In order to extract a period from the \gacf{}, we elected to use a Fast Fourier Transform \citep[FFT;][]{Cooley1969}. 
Extracting periods from an ACF can be done in a number of ways, most simply by selecting the first (or largest) peak in the ACF \citep[e.g. as in][]{McQuillan2014}. This can lead to inaccuracies, in particular for weaker signals as this relies on the first peak being prominent in the ACF. We elected to use an extraction method that relies on the periodicity of the ACF, and the regular sampling of the \gacf{} lends itself to an FFT. Other more complex methods such as fitting a damped harmonic oscillator to the ACF have been used previously \citep{Angus2018}. This in general did not alter extracted periods enough to warrant the additional complexity for such an exploratory work. We also experimented with using fewer ACF peaks rather than the entire signal in order to refine the period, but again the additional complexity was deemed unnecessary for a large scale rotation study.

The FFT is a robust and well-documented method of extracting periodic signals. In this study we used the implementation in the \texttt{numpy.fft} package \citep{Harris2020}. We calculated the FFT with a padding factor of 32, to allow precise resolution of peaks in the Fourier transform. As phase information is lost in taking the ACF of the initial data, a real Fourier transform is sufficient. 

To extract the most likely frequencies, we searched for peaks in the Fourier transform.
{A peak is defined as the central point in a contiguous sequence of 5 points which monotonically increases to the peak, followed by a monotonic decrease from the peak. Additionally, the amplitude of a peak must be greater than 20\% of the highest peak in the periodogram to be included.} Here an automated cut was made - any Fourier transforms with more than 6 peaks were removed as noise. This threshold was selected based on a manual vetting process for one \ngts{} field (10,000 objects) which demonstrated that for these objects with `noisy' Fourier transforms less than 1\% had genuine periodic signals. Removing these objects entirely greatly reduced the number of false positives extracted without removing many `real' signals. 63\% of processed objects were flagged as having no significant periodicity based on this FFT check.

\subsection{Long Term Trend Assessment}
\label{subsec:longtermtrend}
A time baseline of  $\sim 250$ days allows for the extraction of periodic signals up to $\sim 125$ days long. Signals longer than this may be present in the data, however observing one or fewer complete variability cycles cannot definitively characterise a periodic signal. This variability may not be periodic, but rather a long term trend in the data arising from instrumental or telluric changes over these timescales. These objects may still contain interesting periodic variability at a shorter timescale, so by detecting and removing a long term trend we can more accurately calculate the period and amplitude of this variability.

If the most significant peak in the FFT (see Section \ref{subsec:fft}) was at a period greater than half the length of the signal baseline it was flagged as a long term trend. When this occurred we computed a high-pass filter for the signal by calculating the median flux at each time step in a rolling window which is 10\% of the time extent of the light curve. This captures any long term behaviour without removing any shorter period variability.
We divided this median filter from our signal and re-ran the cleaned light curve back through the signal detection pipeline. If no signal of interest was detected at this stage (either we found noise or residuals of our median filter), the object was flagged as having a long term trend and removed from processing.

\subsection{Moon Signal Assessment}
\label{subsec:moon}
During initial testing of the period extraction algorithm, it was noted that a large number of periods between 27 and 30 days were identified by the period search algorithm. Upon closer inspection, these periods had very similar phases and could be split into two groups of signal shapes. The two signal shapes, when phase folded on a new Moon epoch, appeared as a slight increase or decrease in flux at 0.5 phase, i.e. full Moon. This was accompanied by an increase in scatter in the flux measurements at full Moon. Examples of contaminated signals are shown in Figure \ref{fig:moon_data}.

We fitted a model to these Moon correlated noise signals (`Moon signals') and flagged and removed any objects which fitted the expected trend. A detailed description of the model and removal process is given in Appendix \ref{app:moon}.

\subsection{Alias Checks}
\label{subsec:aliases}
As we are using an FFT to extract periodicity from our \gacf{}, we are prone to aliasing. Aliasing is a well known and well-described problem in signal processing, and if the true frequency of the signal and the sampling frequency are known it is trivial to calculate the frequency of aliases as

\begin{equation}
    \label{eq:alias}
    \nu_{\text{alias}} = \nu_{\text{true}} \pm n \cdot \nu_{\text{sampling}}
\end{equation}

\noindent where $n$ is an integer. {We define period as the inverse of frequency, i.e. $P = \frac{1}{\nu}$}.
In the case of ground-based observation, the most common sampling period will be 1 day. In addition, although the background correction should remove this, there will remain residuals of the brightness trend expected throughout the night's observation. Although the sampling of the \gacf{} is regular, the sampling of the inputted light curve will affect the shape of the \gacf{}. We thus expect peaks in the FFT associated with 1-day systematic signals, as well as the true signal aliased with the 1-day sampling.

{For each} light curve we first removed any periods arising from the 1-day sampling. We removed periods within $5\%$ of 1 day, as well as within $5\%$ of integer multiples of 1 day in period and integer multiples of 1 / day in frequency. 
We then assessed whether groups of periods were aliases of one another with respect to common sampling periods using basic graph theory. We construct a graph of frequencies connected by the standard alias formula in Equation \ref{eq:alias}, using sampling periods of one day, 365.25636 days (one year), 27.32158 days (Lunar sidereal period) and 29.53049 days (Lunar synodic period). Each vertex in the graph represents an FFT peak frequency, with connections (edges) made if two frequencies can be related to one another through Equation \ref{eq:alias} given one of our sampling frequencies. Note we considered aliases arising from both the synodic and sidereal Lunar period, however, given the $5\%$ tolerance used for assessing similarity, these two sampling frequencies connected the same frequencies in the majority of examples.

For each connected sub-graph (i.e. a group of frequencies connected by the same sampling aliases) we {determined} the frequency for which the phase folded light curve had the lowest {spread in flux and took this to be the correct period}. We calculated the $5^{\mathrm{th}} - 95^{\mathrm{th}}$ percentile spread in flux within bins of $0.05$ width in phase and then calculated the average of these values weighted by the number of points within each flux bin. In addition to the FFT peak periods, we also checked the RMS of twice and half the periods, as in some cases we found twice the FFT peak period was the correct period. This was assessed by-eye initially, and {appeared to be} much more common for short period objects due to aliasing from the 1-day sampling. This same approach was taken by \citet{McQuillan2013}, however, we elected to automate the process rather than by-eye confirmation.

\subsection{Further Signal Validation}
\label{subsec:field_validation}

Due to the ground-based nature of \ngts{}, some fields were not continuously observed for the entirety of the field time-baseline. {As a result of bad weather and technical downtime, there were gaps in observations lasting several weeks for a number of the fields used in this study.} In these cases it is no longer correct to use the entire time baseline as a cut-off for robust periods. Instead, we elected to find the longest period of continuous observation within these fields and remove any periods greater than half this time length. We define a period of continuous observation as a period in which there are no observation time gaps of greater than $20\%$ of the entire field baseline. For our 250-night observation baseline, this equates to gaps of 50 days or longer. This removed 907 detected periodic signals {from 11 different fields}, and manual inspection of the removed signals confirmed that many of the removed detections were systematic periods arising from the long sampling gaps, rather than astrophysical variability.

Additionally, a number of detected periodic signals with unphysically large amplitudes were detected. On inspection it appears these signals were incorrectly processed by the \ngts{} pipeline, resulting in non-physical flux values. In our final sample, we elected to remove any signals with a relative amplitude $> 1.0$. This removed 58 signals, and manual inspection of the removed signals confirmed the majority of signals removed were non-physical; especially for the largest amplitude signals. The cut-off was chosen empirically based on the signal amplitude distribution of our sample. 

Our initial search resulted in 17,845 periodic detections. Removing 907 long term trends left 16,938 detections. Finally, removing 58 unphysically large amplitude signals resulted in 16,880 detections.

\subsection{Cross-matching with Gaia DR2 \& TICv8}

In order to assess our variability period sample within a meaningful scientific context, we elected to use \gaia{} Data Release 2 \citep[DR2,][]{GaiaCollaboration2018a} for cross-matching and to identify the nature of corresponding objects and their stellar parameters. The \ngts{} database contains cross-matching information with many external catalogues, including \gaia{} DR2. Detail on how the cross-matches are found is given in Section 5 of \citet{Wheatley2018} and briefly in Section \ref{subsec:lc_extraction} of this paper.

As an extension of the \gaia{} DR2 data, the most recent Tess Input Catalogue \citep[TICv8,][]{Stassun2019} contains \gaia{} DR2 data relevant to this study plus additional calculated values and cross-match data. These include more accurate calculated distances from \citet{Bailer-Jones2018} and reddening values which have been used to calculate absolute magnitudes.

More recently, the \gaia{} Early-DR3 \citep{GaiaCollaboration2020} contains improved precision on the astrometric fits to many objects from \gaia{} DR2, however as we are using many derived parameters from external catalogues we elected to continue to use the DR2 parameters throughout this study.

\subsection{Extinction Correction}
\label{subsubsec:extinction}

In the final data products, we assess variability in the context of the colour-magnitude diagram which requires the calculation of absolute magnitudes. In order to be as accurate as possible, we combined \gaia{} G magnitudes ($G$) with distance estimates and accounted for extinction. We used the per-object reddening values from TICv8, multiplied by a total-to-selective extinction ratio of 2.72 to account for the \gaia{} G-band extinction ($A_G$). Further details on how the reddening values and the total-to-selective extinction ratio were calculated can be found in Section 2.3.3 of \citet{Stassun2019}.
Our final value for absolute magnitude was calculated using the formula:
\begin{equation}
M_G = G - 5\log_{10}(\text{distance}) + 5 - A_G.    
\end{equation}

\section{Results}
\label{sec:results}

{Using the G-ACF period extraction pipeline, we derived variability periods for \nperiods{} stars observed with NGTS. A subset of these results is shown in Table \ref{table:catalogue}, along with positions and cross-match data. The format of the results table is shown in Table \ref{table:catalogue_format}.}

\begin{table*}
    \centering
    \caption{Variability periods, amplitudes, positions and catalogue cross-match identifiers for all variable objects in the \ngts{} data set (table format).}
    \begin{tabular}{@{}lllll@{}}
    \toprule
    Column & Format & Units & Label         & Description                                        \\ \midrule
    1      & A18    & ---   & NGTS\_ID      & NGTS source designation                            \\
    2      & F9.5   & deg   & NGTS\_RA      & Source right ascension (J2000)                     \\
    3      & F9.5   & deg   & NGTS\_DEC     & Source declination (J2000)                         \\
    4      & F8.5   & mag   & NGTS\_MAG     & NGTS I-band magnitude                              \\
    5      & F9.5   & days  & PERIOD        & Extracted variability period                       \\
    6      & F7.5   & ---   & AMPLITUDE     & 5-95 percentile relative flux                      \\
    7      & I19    & ---   & GAIA\_DR1\_ID & Cross-matched Gaia DR1 identifier                  \\
    8      & I19    & ---   & GAIA\_DR2\_ID & Cross-matched Gaia DR2 identifier                  \\
    9      & I10    & ---   & TIC\_ID       & Cross-matched Tess Input Catalogue (v8) identifier \\
    10     & A16    & ---   & TWOMASS\_ID   & Cross-matched 2MASS identifier                     \\
    11     & A19    & ---   & WISE\_ID      & Cross-matched WISE identifier                      \\
    12     & A10    & ---   & UCAC4\_ID     & Cross-matched UCAC4 identifier                    
    \end{tabular}
    \label{table:catalogue_format}
\end{table*}

\begin{table*}
    \centering
    \caption{A sample of variability periods, amplitudes, positions and catalogue cross-match identifiers in the \ngts{} data set. A number of catalogue cross-match columns have been excluded for publication clarity. The full table will is available at CDS via anonymous ftp to \url{cdsarc.u-strasbg.fr} (130.79.128.5) or via \url{https://cdsarc.unistra.fr/viz-bin/cat/J/MNRAS}, or as supplementary material.}
    \begin{tabular}{lrrrrrrr}
\toprule
          NGTS ID &  NGTS RA &  NGTS Dec & NGTS Mag &    Period & Amplitude &          Gaia DR2 ID &   TICv8 ID \\
\midrule
  NG0613-3633\_231 & 94.88721 & -35.20762 & 14.77188 & 117.30427 &   0.07218 &  2885392740653834368 &  124854845 \\
  NG0613-3633\_234 & 91.91176 & -35.20084 & 15.86231 & 128.42220 &   0.18731 &  2885953869540806656 &  201389809 \\
  NG0613-3633\_235 & 94.93884 & -35.20675 & 12.91320 & 117.53460 &   0.04857 &  2885392878092780544 &  124854842 \\
  NG0613-3633\_262 & 94.95269 & -35.20598 & 14.51873 & 109.77205 &   0.11461 &  2885392225257749760 &  124854841 \\
  NG0613-3633\_481 & 93.77213 & -35.22205 & 13.69049 &   0.29365 &   0.13175 &  2885521658392050944 &  124689517 \\
  NG0613-3633\_598 & 93.31896 & -35.22787 & 11.48757 &  92.88398 &   0.00860 &  2885530999944081792 &  201530507 \\
  NG0613-3633\_773 & 95.01907 & -35.23832 & 12.55225 & 110.36974 &   0.07016 &  2885380160692365824 &  124855736 \\
 NG0613-3633\_1101 & 95.06110 & -35.25333 & 15.16207 & 128.42220 &   0.22943 &  2885381333220681216 &  124855723 \\
 NG0613-3633\_1181 & 95.06864 & -35.25766 & 13.60488 & 100.74969 &   0.08537 &  2885380577306436736 &  124855720 \\
 NG0613-3633\_1479 & 95.12023 & -35.27187 & 14.86311 & 100.46635 &   0.25487 &  2885380439867479040 &  124922604 \\
\bottomrule
\end{tabular}
    \label{table:catalogue}
\end{table*}

\begin{table}
    \centering
    \caption{A table of the output states of the \nobjs{} \ngts{} objects analysed by the signal detection pipeline. Note a further 907 objects were removed due to large observation gaps in a number of fields, and an additional 58 with spuriously large amplitudes resulting in a final total of \nperiods{} variability periods (see Section \ref{subsec:field_validation}).}
    \begin{tabular}{@{}lcccl@{}}
    \toprule
    Output State    & \multicolumn{1}{l}{Count} & \multicolumn{1}{l}{\% of total} & \multicolumn{1}{l}{\% of detections} &  \\ \midrule
    Bad Data        & 43,358                    & 5.227                           & -                                    &  \\
    Noisy FFT       & 528,105                   & 63.667                         & -                                    &  \\
    Moon            & 175,565                   & 21.166                          & 67.043                               &  \\
    Alias           & 57                        & 0.007                           & 0.022                                &  \\
    Long Term Trend & 64,551                    & 7.782                           & 25.018                               &  \\
    Periodic Signal & 17,845                     & 2.151                           & 6.916                                &  \\ \bottomrule
    \end{tabular}
    \label{table:outputs}
\end{table}

\begin{figure}
    \centering
    \begin{subfigure}[b]{0.9\linewidth}
         \centering
         \includegraphics[width=\linewidth]{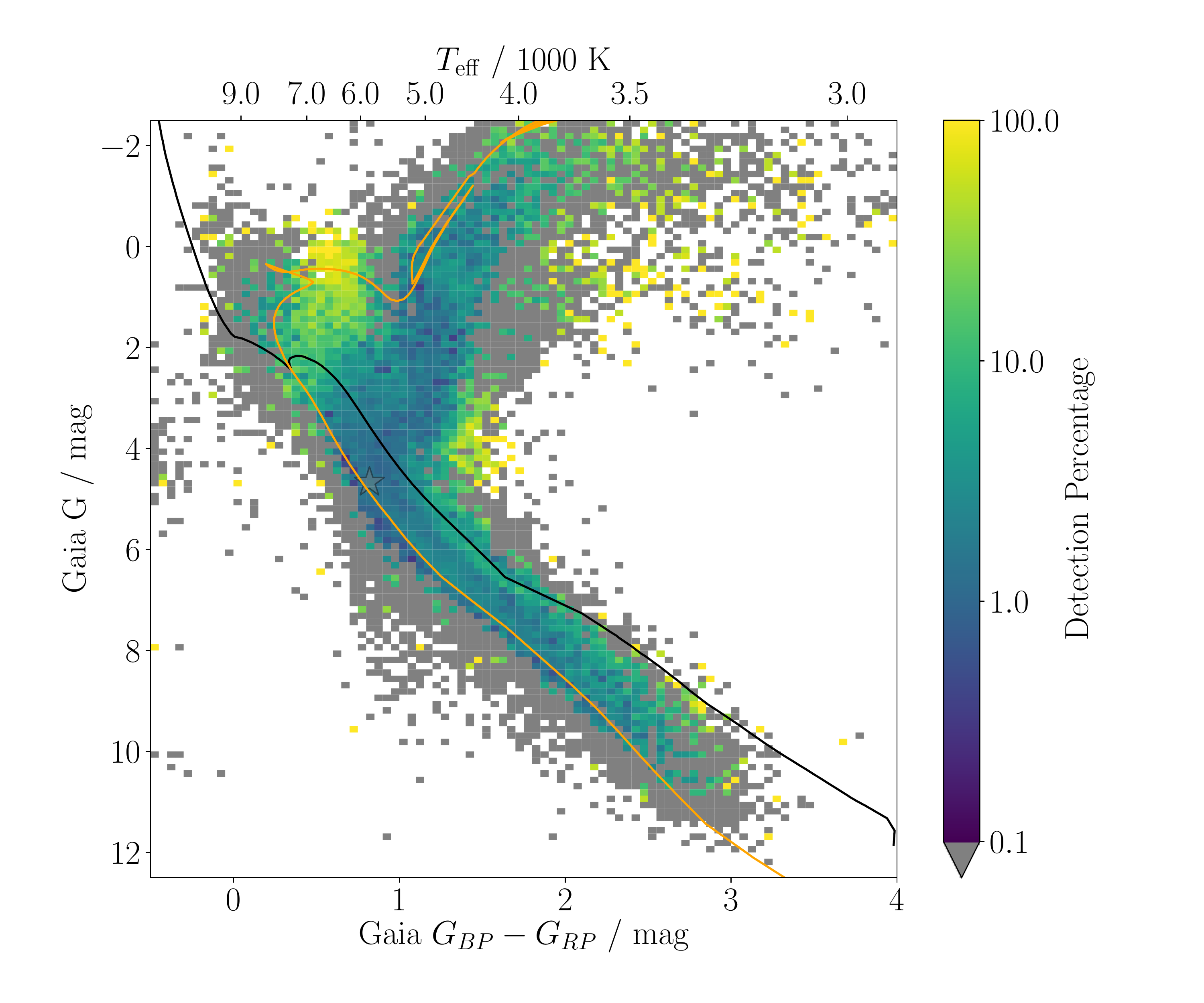}
         \caption{The empirical detection percentage per bin. This is defined as the ratio of the number of detected periodic signals to all observed objects per bin. 0 detections within bins are coloured grey.}
         \label{fig:HR_detection_percentage}
    \end{subfigure}
    \hfill
    \begin{subfigure}[b]{0.9\linewidth}
         \centering
         \includegraphics[width=\linewidth]{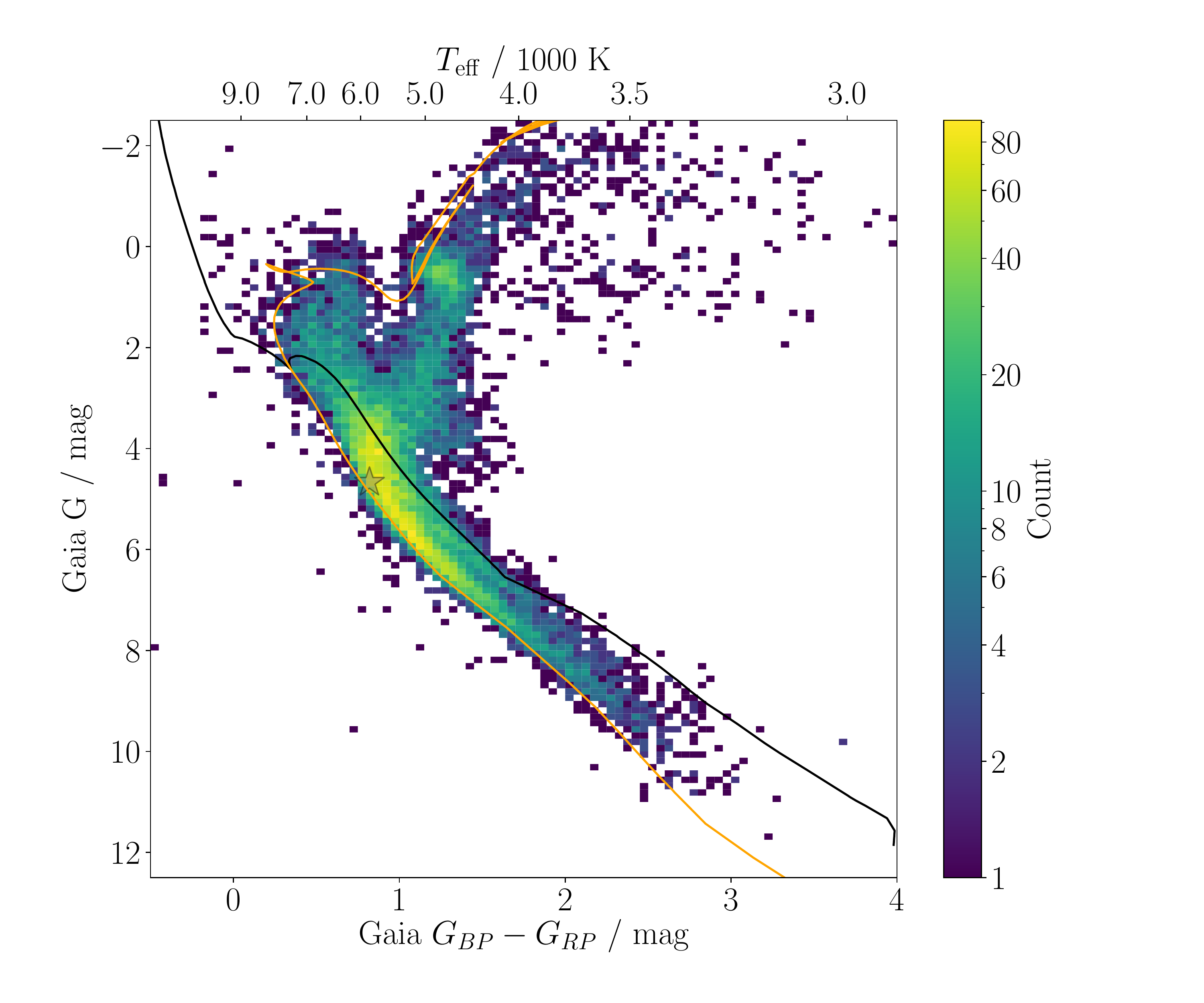} 
         \caption{The number of objects with detected variability within each colour-magnitude bin.}
         \label{fig:HR_count}
    \end{subfigure}
    \hfill
    \begin{subfigure}[b]{0.9\linewidth}
         \centering
         \includegraphics[width=\linewidth]{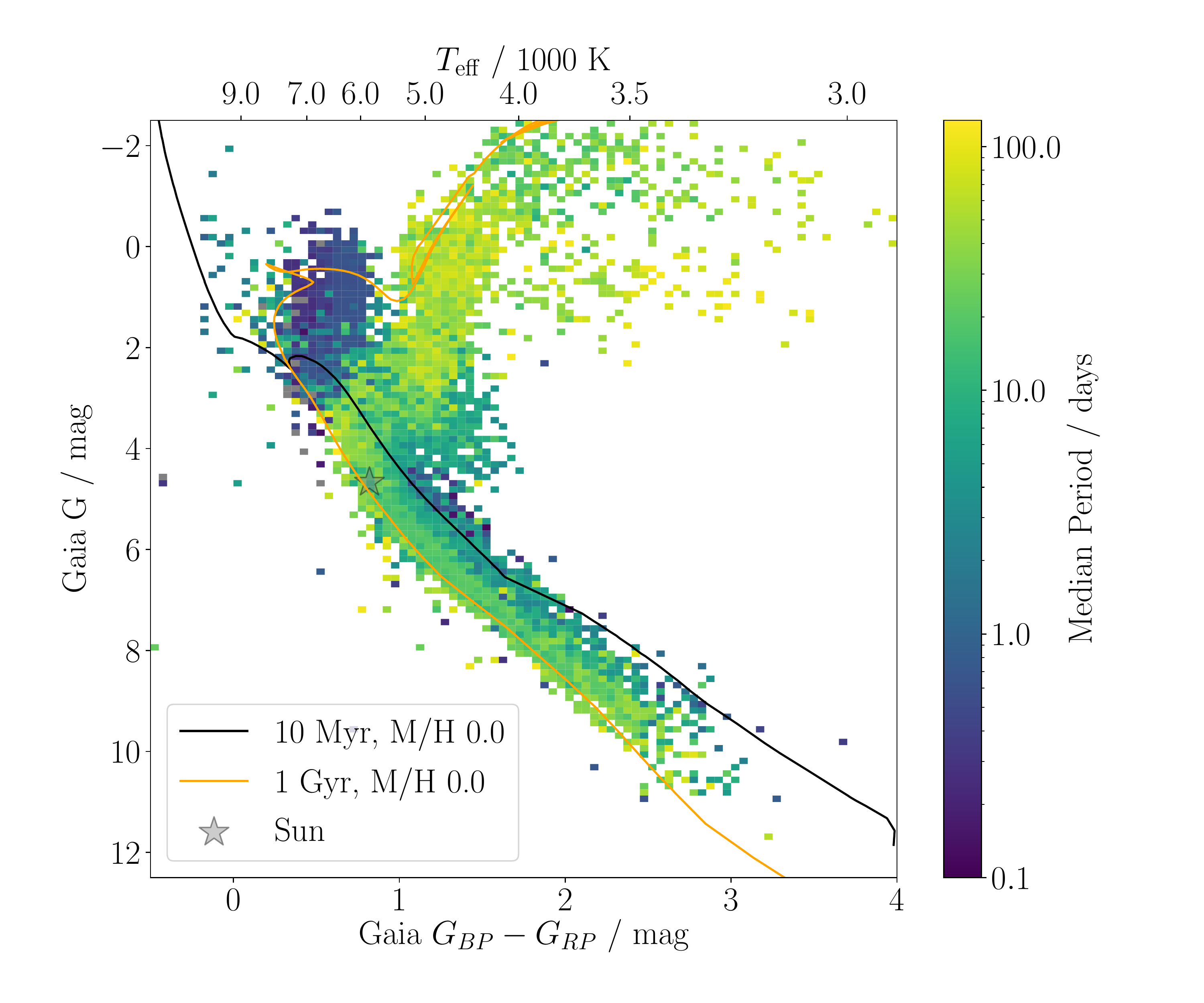}
         \caption{The median variability period within each colour-magnitude bin.}
         \label{fig:HR_median_period}
    \end{subfigure}
    \caption{Binned colour-magnitude (HR) diagrams of the \ngts{} variability sample. PARSEC v1.2 \citep{Bressan2012} Solar metallicity isochrones of ages 10 Myr and 1 Gyr are included as solid black and orange lines, respectively.}
    \label{fig:HR_diagrams}
\end{figure}

\subsection{Periodicity in Colour-Magnitude Space}

Figure \ref{fig:HR_diagrams} shows our variability sample in colour-magnitude space, commonly known as a Hertzsprung-Russell (HR) Diagram {or a colour-magnitude diagram (CMD)}. Table \ref{table:outputs} details the breakdown of outputs from the pipeline. Once cross-matched with TICv8, we were left with a total of \nperiods{} variable light curves from the initial sample of \nobjs{} light curves. This gives a final detection percentage of $2.04\%$. The detection percentage varies in colour-magnitude space as shown in Figure \ref{fig:HR_detection_percentage}, highlighting potential regions of increased variability or increased sensitivity of \ngts{} and the signal detection pipeline.

All conversions between \teff{}, \bprp{} and \grp{} in the following sections are calculated using relations defined in the `Modern Mean Dwarf
Stellar Colour and Effective Temperature Sequence' \citep{Pecaut2013}\footnote{A more recent version of the table including Gaia DR2 colours is maintained at \url{http://www.pas.rochester.edu/~emamajek/EEM_dwarf_UBVIJHK_colors_Teff.txt}}, interpolated using a univariate cubic spline.
The isochrones in the HR diagrams are taken from PARSEC v1.2S \citep{Bressan2012}. We elected to use these isochrones as they have been proven to fit the \gaia{} DR2 main sequence well in \citet{GaiaCollaboration2018b}. We produce isochrones using PARSEC v1.2S, selecting the \gaia{} DR2 passbands from \citet{Evans2018}\footnote{using the CMD 3.4 input form at \url{http://stev.oapd.inaf.it/cgi-bin/cmd}}. The isochrone at 1 Gyr gives a good indication of where the main sequence lies, with the earlier age isochrone at 10 Myr indicating locations on the HR diagram of potentially younger stellar populations. We note, as shown in \citet{Gillen2020b}, that the PARSEC v1.2 models appear to be less reliable at pre-main-sequence ages, but should be sufficient for their indicative use in this study.

Figure \ref{fig:HR_detection_percentage} highlights regions of interest in terms of detection percentage. Additionally, Figure \ref{fig:HR_count} shows the number of detections in each bin. Where detection percentage approaches 100\% this is often indicative of a single variable object falling in this colour-magnitude bin. As in \citet{GaiaCollaboration2019}, we identify distinct regions of variability within the HR diagram and suggest the types of variable objects which may lie at each location.

The region at the top of the main sequence (\bprp{} $\sim0.4$, $G \sim1.0$) reveals a high proportion of variable objects.
We also see a region of increased variability at the `elbow' of the main sequence and the Red-Giant Branch (RGB) (\bprp{}$\sim1.5$, $G \sim4$). These objects may be young, massive objects with high levels of activity.

In Figure \ref{fig:HR_median_period} we plot the median period in each colour-magnitude bin. Of particular interest, we see distinct regions of different variability periods on the HR diagram. 
There is a region of short median period at the top of the main sequence (\bprp{} $\sim0.4$, $G \sim1.0$).
{Typical spot-driven photometric modulation will not be present on these hotter, radiative stars. The majority of variability seen in this region likely arises from pulsations. There may also be a number of magnetic OBA or chemically peculiar Ap stars within this region. In these stars, photometric brightness fluctuations are seen as a result of fossil magnetic fields imprinting chemical abundance inhomogeneity on the stellar surface \citep{Sikora2019, David-Uraz2019}. These targets are prime candidates for future spectropolarimetric observations \citep[e.g.][]{Grunhut2017}.}

A large number of the longest period variability signals lie on the RGB (\bprp{} $\gtrapprox1.0$, $G \lessapprox2.0$. These signals could indicate extremely slowly rotating large stars or other photometrically varying sources such as giant star pulsations.

We also see a clear trend of increasing period as we move perpendicular down towards the main sequence along the Hayashi tracks \citep{Hayashi1961}. There are potentially a number of effects at play here:
\begin{itemize}
    \item[1)] We would expect a population of equal mass binary stars with short rotation periods to lie $0.75$ in absolute magnitude above the main sequence, contributing to the shorter median period in this range.
    \item[2)] We would also expect a population of young stars to lie in this region of colour-magnitude space. In particular, we see short period objects which lie between the 10 Myr and 1 Gyr isochrones. 
\end{itemize}
In this region of the HR diagram potentially lie pre-main-sequence (PMS) Young Stellar Objects (YSO) such as T-Tauri stars with protostellar debris discs, which we expect to have shorter rotation periods than main-sequence stars {of the same mass (colour)}.
The median period observed for the bulk of main-sequence objects is 20 to 30 days, as expected.

\begin{figure}
    \centering
    \includegraphics[width=\linewidth]{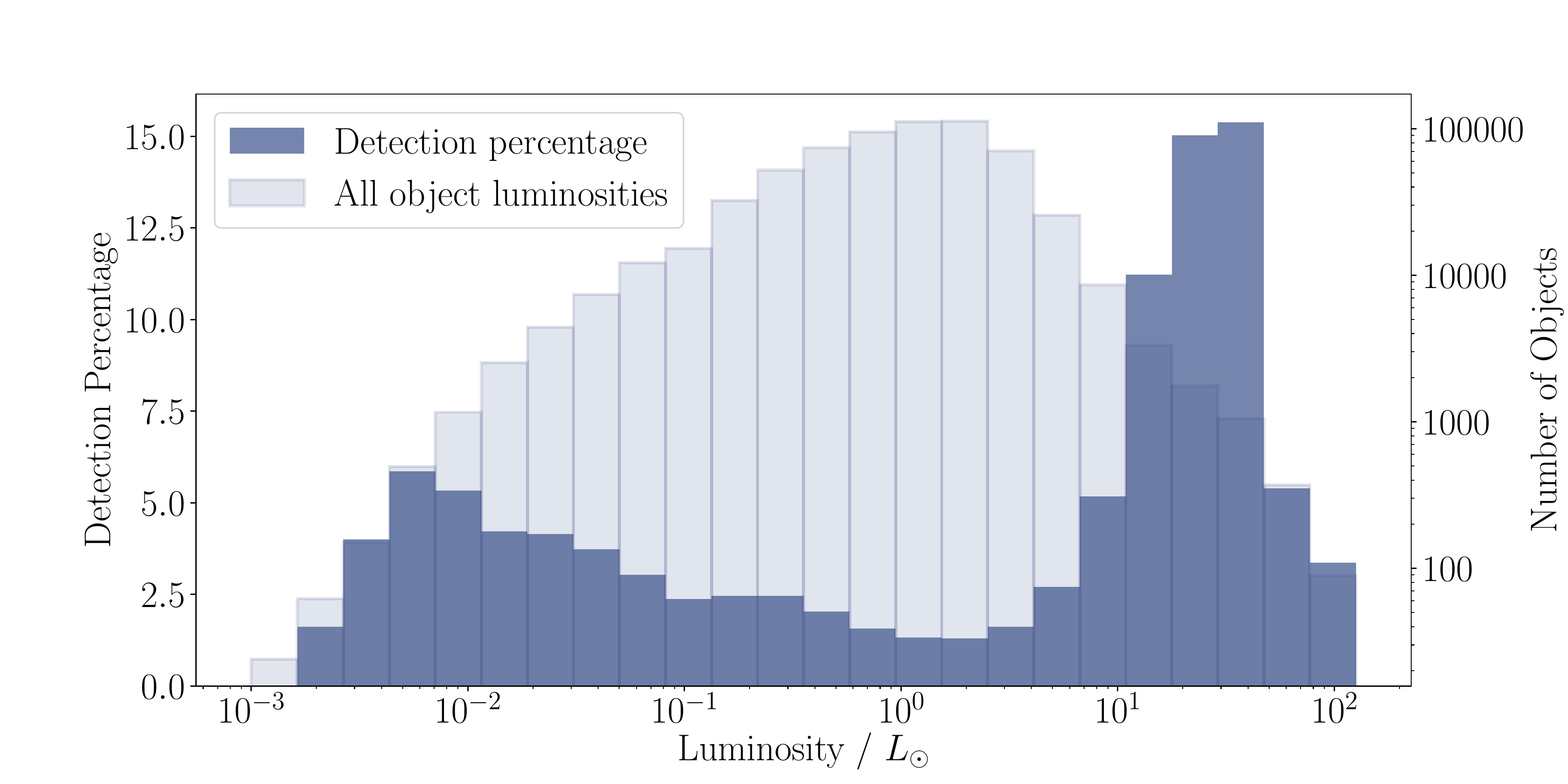}
    \caption{Histogram of the empirical detection percentage {(left y-axis)} for all sources against luminosity, as well as the luminosity distribution for all observations {(right y-axis)}.}
    \label{fig:luminosity_detection_percentage}
\end{figure}

We plot detection percentage vs. luminosity in Figure \ref{fig:luminosity_detection_percentage}. Luminosity values are taken from TICv8 \citep{Stassun2019}, calculated using Equation \ref{eqn:lum}.

\begin{center}
\begin{align}
    \frac{L}{L_{\odot}} = \left(\frac{R}{R_{\odot}}\right)^2 \cdot \left(\frac{\text{\teff{}}}{5772}\right)^4
    \label{eqn:lum}
\end{align}
\end{center}

We use the radii values provided by TICv8. These radii values are either taken from pre-existing dwarf catalogue values {\citep[from ][]{Muirhead2018}}, or when these are not available (as is the case for a large majority of the \ngts{} sources) they are calculated from distance, bolometric corrections, G magnitude and a preferred temperature. Full details of this calculation are given in \citet{Stassun2018}. \teff{} values come from spectroscopic catalogues where available, otherwise they are derived from the de-reddened \bprp{} colour.

As expected, we recover a much higher fraction of variable signals from more luminous stars, with {up to} 15\% of the most luminous objects in our sample having detectable variability signals. These objects will correspond to luminous giant stars, where we would expect large-amplitude variability arising from pulsations.
The lowest number of variable objects coincides with the peak in the number of objects {(at  1.5--2.5 $L_{\odot}$)}, where we detect variability in $<2\%$ of objects.
We also observe an increase in detection percentage for the faintest objects. Here we should expect to be observing cooler dwarf stars and young stars which generally have higher levels of magnetic activity and could lead to increased detection of photometric variability. 
Additionally, close binaries may appear more luminous than single stars and from their position above the main sequence in the HR diagram (Figure \ref{fig:HR_diagrams} (a)) {appear to} have a higher detection percentage {than equivalent single stars}. Given the width of the luminosity bins used {is larger than the expected luminosity increase from a single star to an equal luminosity binary} ($0.2$ dex, a factor $\sim 1.6$ in luminosity), this will not have a large effect on the plotted distribution.

{We assessed the distribution of detection percentage against on-sky RA and Dec for our population.} The distribution of detection percentage {for field stars} did not appear to have any obvious correlation with {on-sky} position.
    
\subsection{Example Variability Signals}

We show six examples of variability signals in Figure \ref{fig:example_lcs}. A table of stellar parameters for each object is included for reference. 
We selected the included objects to demonstrate a small selection of the variability we are able to extract from \ngts{} light curves. The stars are selected to have a range of spectral types, and demonstrate variability with different periods, amplitudes and signal shapes. In particular, using the object numbering as in Figure \ref{fig:example_lcs} (1 to 6, top to bottom):

\begin{enumerate}
    \item[1)] An extremely short period, semi-detached eclipsing binary. This object lies above the main sequence, as expected for a near-equal mass binary system. 
    \item[2)] A typical short period pulsation signal from an RR-Lyrae object.
    \item[3)] A candidate young stellar object (YSO). Objects above the main sequence with periods of 1 to 10 days are excellent YSO candidates, suitable for follow-up infrared and spectroscopic observations.
    \item[4)] {An example of a variable red-giant star. These are stars such as Cepheids, semi-regular variables, slow irregular variables or small-amplitude red-giants.}
    \item[5)] A main-sequence late-G dwarf star, with small amplitude 20- to 30-day variability.
    \item[6)] A long period M-dwarf.
\end{enumerate}

Within the observed \gacf{} signals we see artefacts arising from 1-day sampling aliases. These aliases are particularly relevant for signals of period $< 1$ day, where it was necessary to perform the additional verification steps outlined in Section \ref{subsec:aliases}.

\newcommand{\myfigsize}{0.8}
\begin{figure*}
    \includegraphics[width=0.3\linewidth]{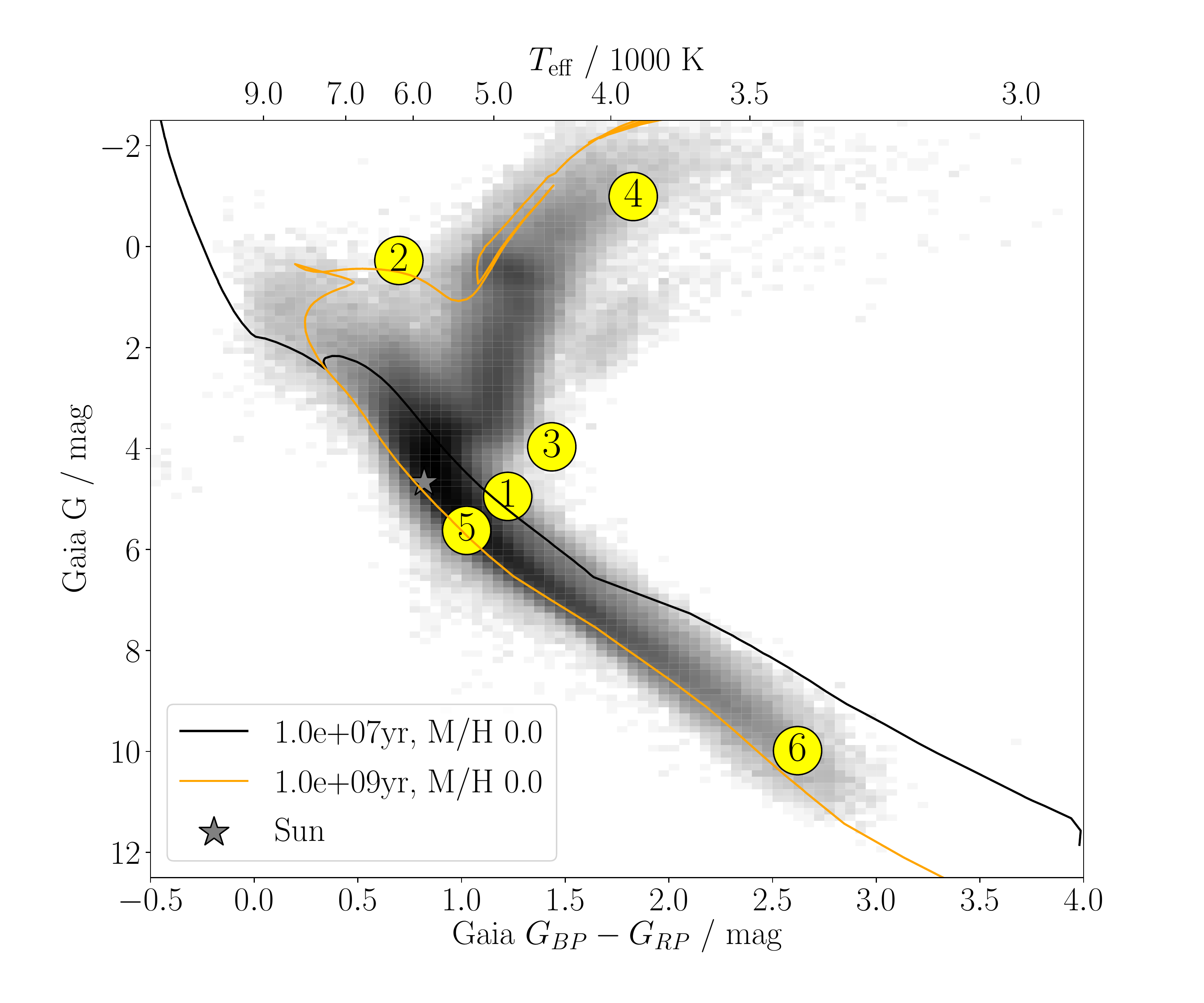} \par
    
    (1) \raisebox{-.5\height}{\includegraphics[width=\myfigsize\linewidth]{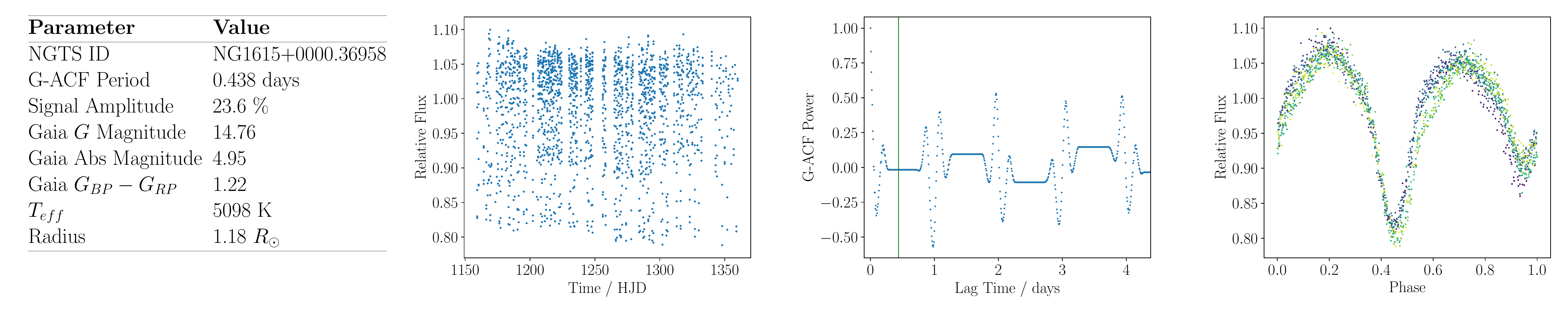}}
    
    (2) \raisebox{-.5\height}{\includegraphics[width=\myfigsize\linewidth]{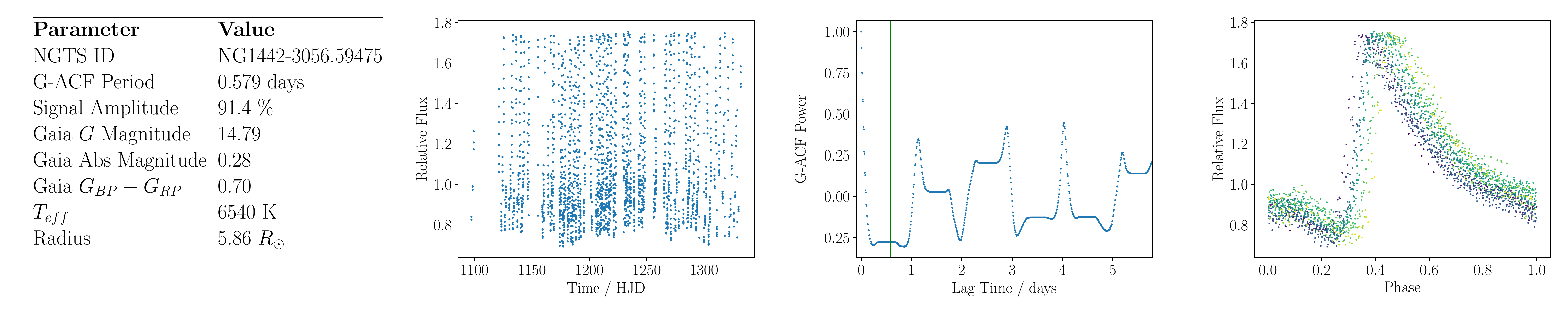}}
    
    (3) \raisebox{-.5\height}{\includegraphics[width=\myfigsize\linewidth]{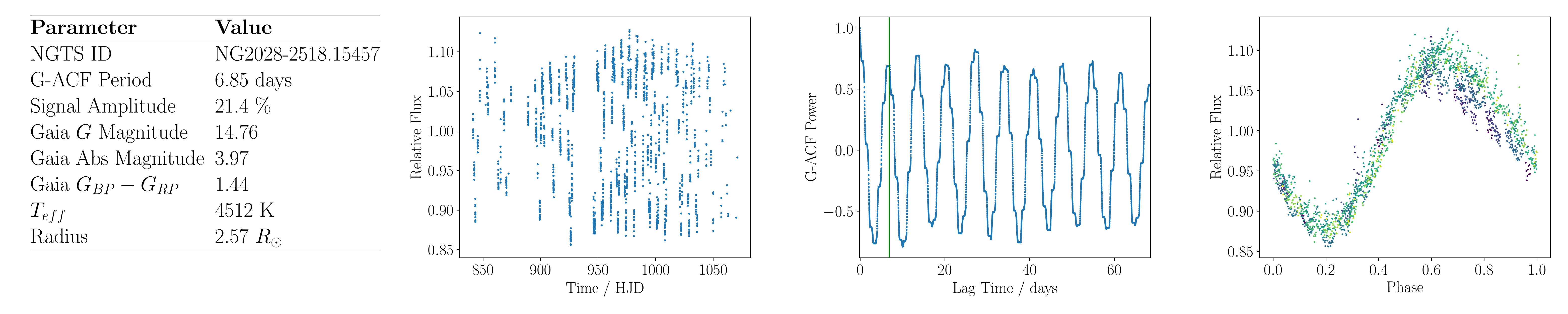}}
    
    (4) \raisebox{-.5\height}{\includegraphics[width=\myfigsize\linewidth]{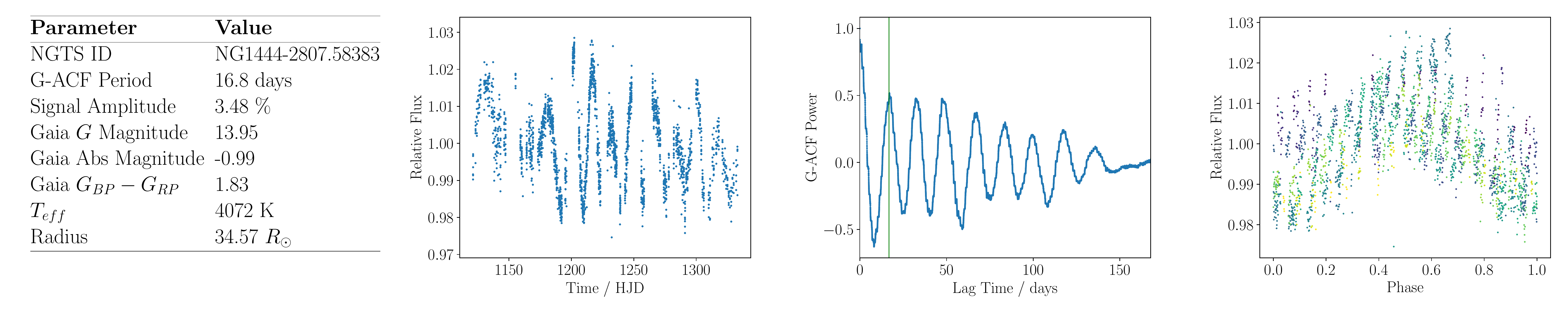}}

    (5) \raisebox{-.5\height}{\includegraphics[width=\myfigsize\linewidth]{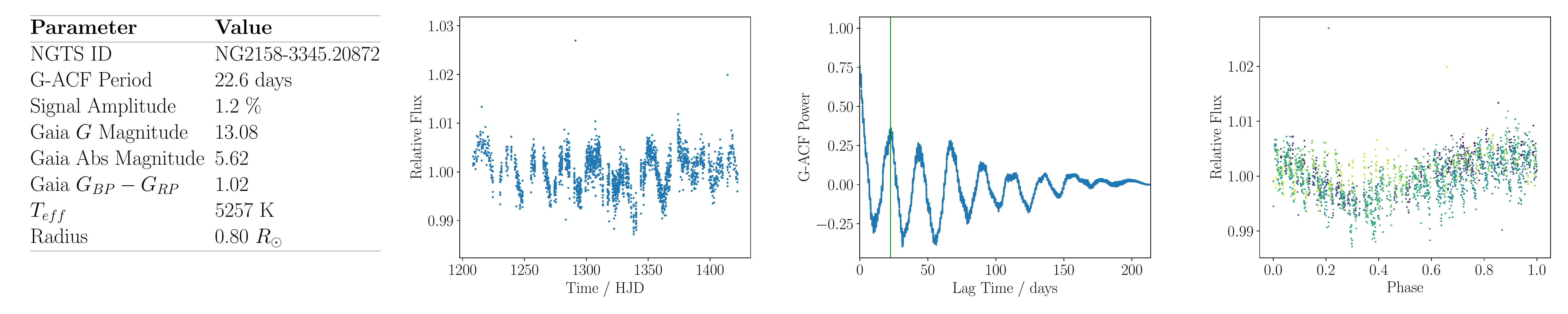}}
    
    (6) \raisebox{-.5\height}{\includegraphics[width=\myfigsize\linewidth]{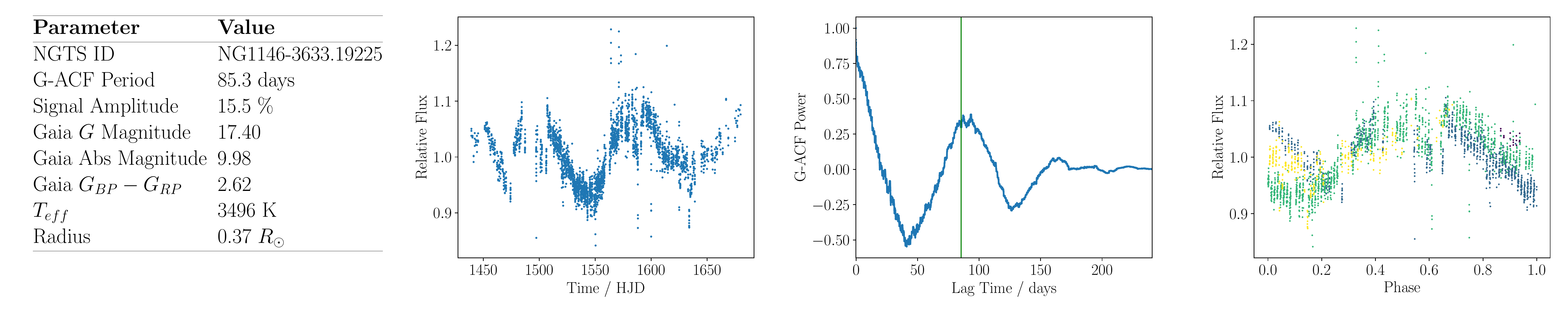}}
	\caption{Example variable star signals across the HR diagram. From left to right: A table of stellar parameters. The \ngts{} light curve, binned to 20 minutes. The \gacf{} of the light curve. The light curve phase folded on the extracted period, each successive period is coloured according to a perceptually uniform sequential colourmap.
	\newline
	The position of each star on the HR diagram is shown, the numbered labels 1 to 6 correspond to the stars top to bottom. Solar metallicity PARSEC isochrones of ages 10 Myr and 1 Gyr are included as solid black and orange line on the HR diagram, respectively.}
	\label{fig:example_lcs} 
\end{figure*}

\subsection{Cross-matching with previous catalogues}

\begin{figure*}
    \centering
    \includegraphics[width=0.45\textwidth]{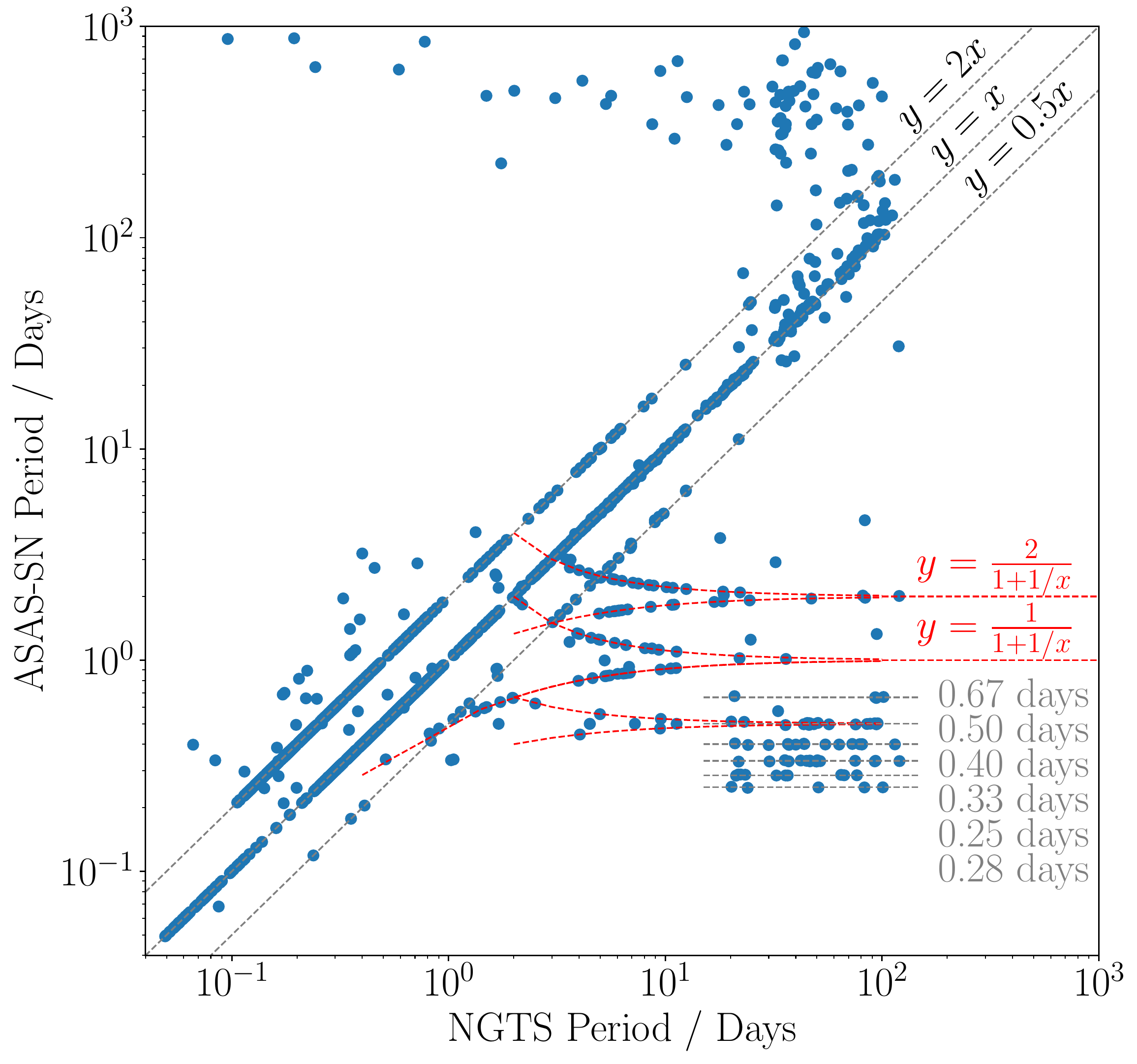}
    \includegraphics[width=0.45\textwidth]{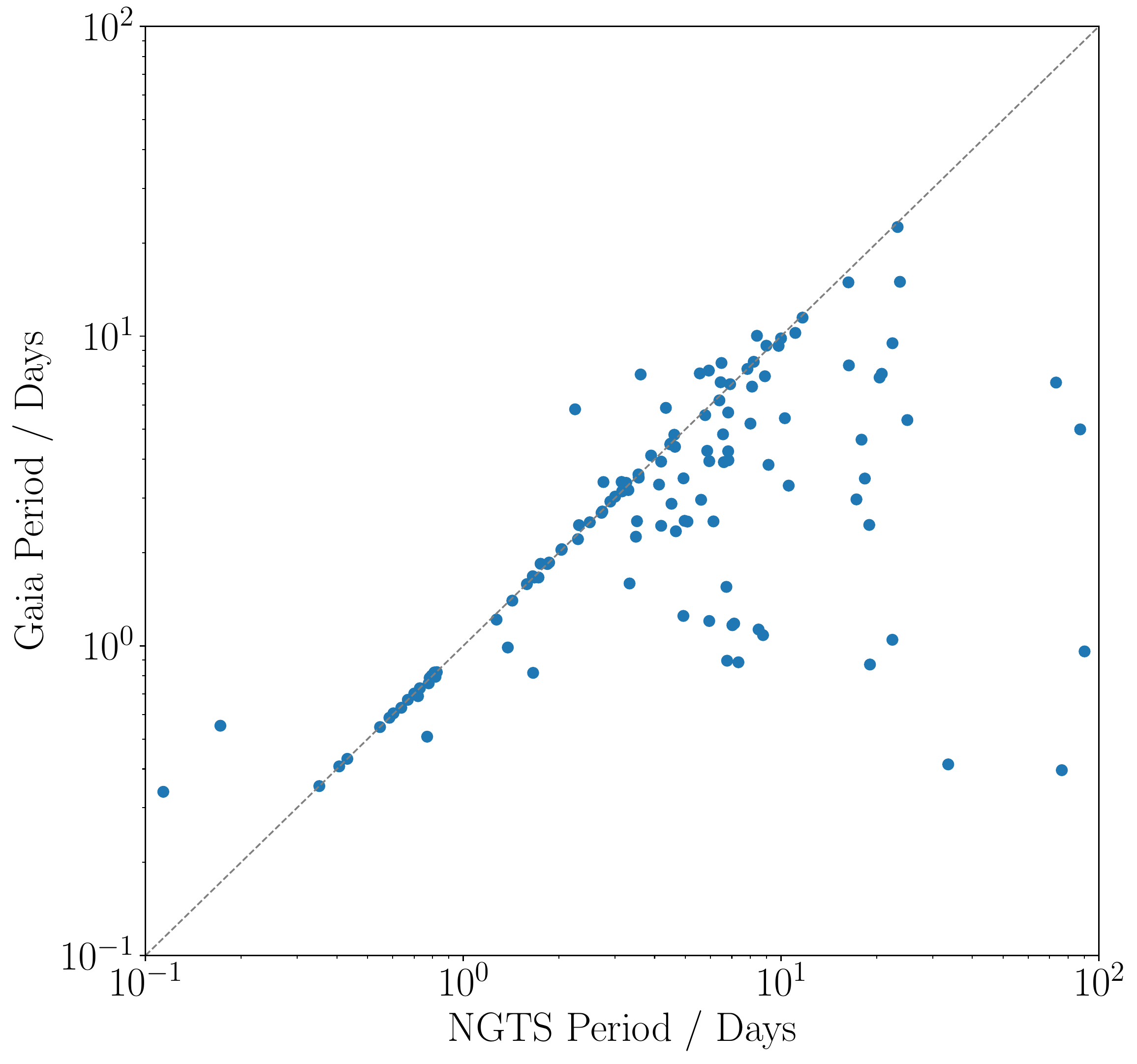}
    \caption{{NGTS variability periods from this study compared with ASAS-SN periods (left) and Gaia (right). Lines of equal period from both surveys are plotted in light grey, and for the ASAS-SN comparison lines showing periods differing due to incorrect phase folding by a factor two shorter or longer are also plotted in light grey. The red dashed lines and associated equations indicate relations between periods arising from 1-day sampling. Light grey dotted horizontal lines in the left-hand figure and corresponding periods indicate where ASAS-SN has recovered periods corresponding to exact fractions of a day.}}
    \label{fig:survey_comparison}
\end{figure*}

{We cross-matched our NGTS variability periods with photometric variability catalogues in the literature.
The ASAS-SN variability catalogue is a large catalogue of photometric variability. We took the latest available data, containing 687,695 variable stars from \citet{Jayasinghe2018} through to \citet{Jayasinghe2021}\footnote{The full catalogue is available at \url{https://asas-sn.osu.edu/variables}}. We cross-matched our catalogue with the ASAS-SN catalogue, matching on TICv8 ID and Gaia DR2 ID. We found 2,439 matches with periods in both catalogues. A period-period comparison is shown in the left panel of Figure \ref{fig:survey_comparison}.
The majority (about 1,500 stars) had similar periods from both catalogues. For approximately 750 stars, the periods differed by a factor of 2. This was most common for eclipsing binary targets in which the primary and secondary eclipses were of similar depths, and either the NGTS or ASAS-SN period was half the correct period. Periods with large discrepancies appear to be long term trends within the NGTS or the ASAS-SN data masking any shorter-term variability, or period aliasing resulting from the 1-day sampling seen in both surveys. The NGTS period extraction pipeline will not return periods close to 1 day or multiples thereof to reduce the number of systematic false positive detections. We see a number of periods in the ASAS-SN catalogue falling on exact fractions of 1 day, resulting in the `stripes' of periods seen in the lower right of the Figure.
We see structures within the period-period diagram resulting from objects for which the NGTS and ASAS-SN detections are aliases of one another with respect to 1-day sampling. Equation \ref{eq:alias} can be used to calculate these connections and relations of the form
\begin{equation}
    P_{\text{ASASSN}} = \frac{1}{P_{\text{sampling}} \pm \frac{1}{P_{\text{NGTS}}}}
\end{equation}
are shown in Figure \ref{fig:survey_comparison}. Three obvious sets of aliased periods exist which trace these relations, accounting for approximately 114 matches. We see two sets of related periods arising from 1-day sampling, with the same double phase folding for eclipsing binaries resulting in the set of periods approaching 2 days. There is also a small group of periods connected by aliases arising from 2-day sampling, however, the form of the relation is not shown in the Figure.}

{We were able to find three cross-matches with the MEarth rotation catalogue from \citet{Newton2018}. Of these, NGTS was able to extract a short 0.4 day rotation period for an object which not present in the MEarth catalogue (NG1444-2807.12982). The two other objects (NG1214-3922.6732 \& NG0458-3916.13434), NGTS detected a near 100 day period, similar to MEarth. The length of these periods would require extended observation from either survey to improve the accuracy as both surveys were only able to observe two to three complete variability cycles.}

{A variability study was conducted as part of the Gaia Data Release 2 \citep[DR2,][]{GaiaCollaboration2018a}, where photometric time-series data was processed to detect and classify variable sources \citep[as described in][]{Holl2018}. Photometric time series from Gaia are sparsely sampled and not optimised to detect photometric variability, so may produce an incorrect period. We cross-matched 126 objects against the rotation period database provided by the Gaia Collaboration on VizieR\footnote{\url{https://vizier.cds.unistra.fr/viz-bin/VizieR-3?-source=I/345/rm}}, these period comparisons are shown in the right panel of Figure \ref{fig:survey_comparison}. For 60 of the 126 periods that differed, we phase folded the NGTS data on both periods and manually inspected which phase fold appeared to be favourable. The NGTS period was favoured in the majority of cases through visual inspection. As expected for space-based data we do not see any aliasing artefacts in the Gaia periods as in the cross-matching with ASAS-SN. This is a clear demonstration that the NGTS period recovery pipeline is well suited to deal with aliases arising from 1-day sampling}

Finally, we cross-matched our sample with the variability catalogue from \citet{CantoMartins2020}, which searched for rotation periods in 1000 TESS objects of interest. We found six objects in both catalogues by matching on TIC id. These come from three different results tables from \citet{CantoMartins2020}: TIC 14165625 and 77951245 contain `unambiguous rotation periods', TIC 100608026 and 1528696 contain `dubious rotation periods' and TIC 150151262 and 306996324 contained no significant variability in the TESS data.
Manual inspection of these objects confirmed the NGTS light curves contained variability at the reported period from this study. For TIC 14165625, the reported TESS period was approximately half the NGTS period, and for TIC 77951245 the reported periods were similar (5.8 days and 5.4 days for NGTS and TESS, respectively), although the phase fold on the NGTS data was cleaner using the NGTS period.

{Although a large number of photometric variable stars are known in the Kepler field, we are unable to cross-match with these catalogues as we do not observe this part of the sky. Additionally, we do not attempt to cross-match with small catalogues and papers reporting detections of individual variable objects. 
Two large variability catalogues we do not attempt cross-matches with are The Zwicky Transient Facility (ZTF) catalogue of periodic variable stars \citep{Chen2020} or the catalogue of variable stars measured by the Asteroid Terrestrial-impact Last Alert System (ATLAS) \citep{Heinze2018}.  The ZTF catalogue contains 4.7 million candidate variables and the ATLAS catalogue 621,702 candidate variables. Both surveys target much fainter objects than NGTS: the brightest candidates in both surveys are approximately as bright as the faintest objects observed by NGTS \citep{Masci2019, Tonry2018}. Due to the small overlap in brightness and a large number of candidates in each catalogue, we elected not to perform a cross-match.
Further cross-matching with smaller catalogues is possible, as we provide the position in RA and Dec, as well as TICv8 and Gaia DR2 identifiers (where available) for all \nperiods{} variable sources.}

\subsection{Period ranges of interest}
\begin{figure*}
    \includegraphics[height=0.7\paperheight]{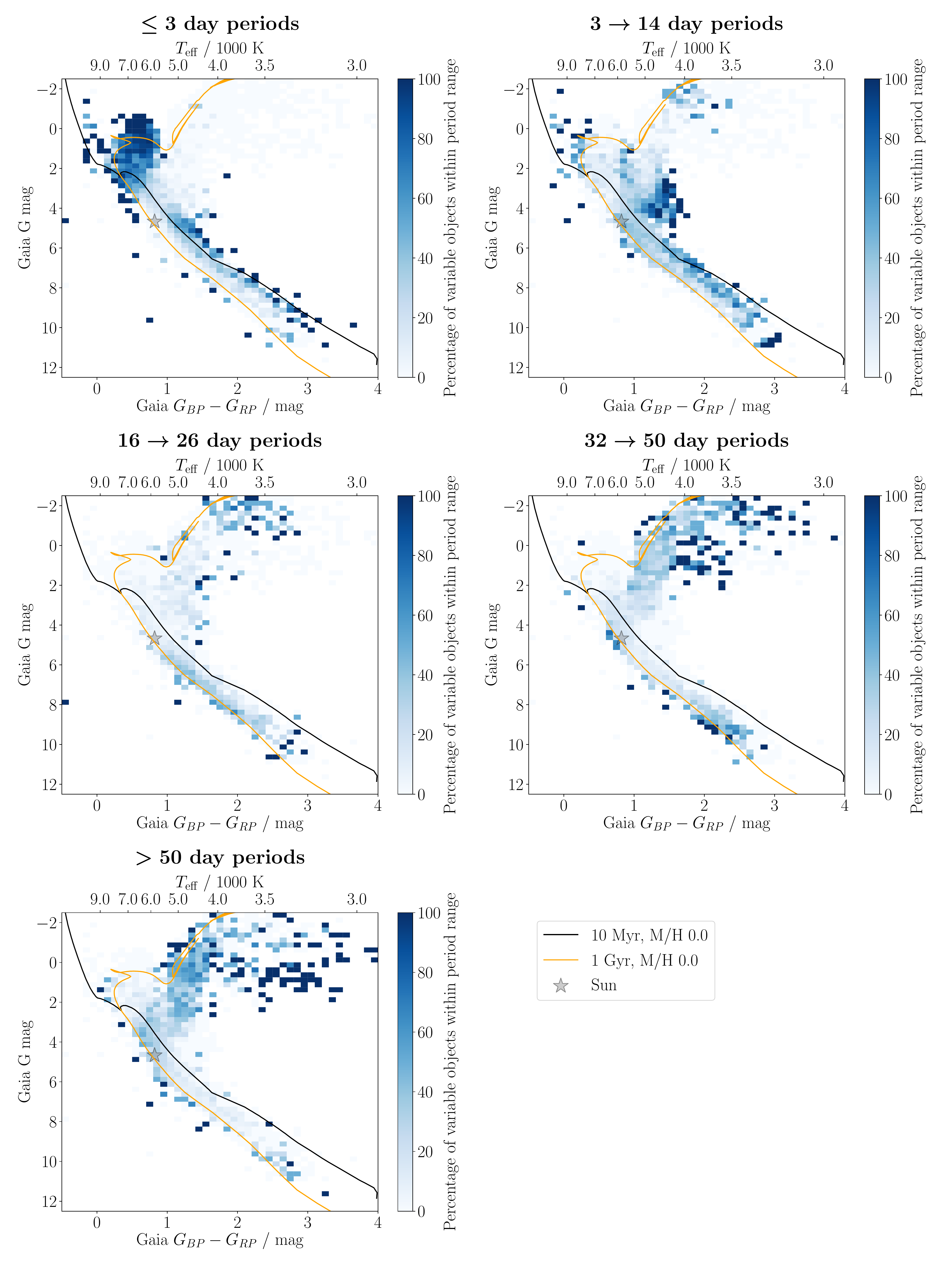}
    \caption{HR diagrams for the \ngts{} variability sample broken down into period ranges. {Periods in the sample range from $\sim$ 0.1 to 130 days.} The colour bar indicates the percentage of all variable objects across all period ranges which lie in this specific colour-magnitude-period bin. The sum of each bin across all 5 subplots will equal 100\%. Solar metallicity PARSEC isochrones of ages 10 Myr and 1 Gyr are included as solid black and orange lines, respectively.}
    \label{fig:HR_diagram_period_breakdown}
\end{figure*}

We break our results down into unevenly spaced intervals in variability period in order to assess how samples of similar variability periods are distributed in colour-magnitude space in Figure \ref{fig:HR_diagram_period_breakdown}. This reveals more information than Figure \ref{fig:HR_diagrams} as we are able to probe into the high-density main sequence. We have selected the period ranges empirically taking into account the sampling gaps at 14 and 28 days arising from Moon contaminated signals.

The majority of the shortest period variability lies at the top of the main sequence. This could be indicative of $\delta$-Scuti, RR-Lyrae or rapidly oscillating Ap stars in the instability strip. Typically, RR-Lyrae type objects lie in this region at the lower end of the instability strip and pulsate with periods of less than 1 day.
The peak density for less evolved stars is above the main sequence at this period range.
Between {1} and 10 days, we would expect to observe the rotation of YSOs such as T-Tauri stars or young main-sequence stars \citep[e.g. as seen in ][]{GaiaCollaboration2019}. 
We may also observe short-period binary star systems at this period range, which would also lie above the main sequence on the HR diagram.
In the period range 3 to 14 days we continue to see a peak density above the main sequence, though the bulk moves towards later spectral types compared to the very short periods.

Between 16 and 26 days, we see the peak density move towards the main sequence as well as a distinct lack of objects above the main sequence.
At $>30$ days, we start to see detections into the RGB as well as more M-type stars. We would expect giant, evolved stars to have longer period rotation or pulsations.
Moving from between 32 and 50 day to $>50$ day periods we see the bulk of objects move further up the RGB and down the main sequence {towards cooler temperatures and redder colours.}

\subsection{Period-colour distribution}
\label{subsec:period-colour}

\begin{figure*}
    \includegraphics[width=\textwidth]{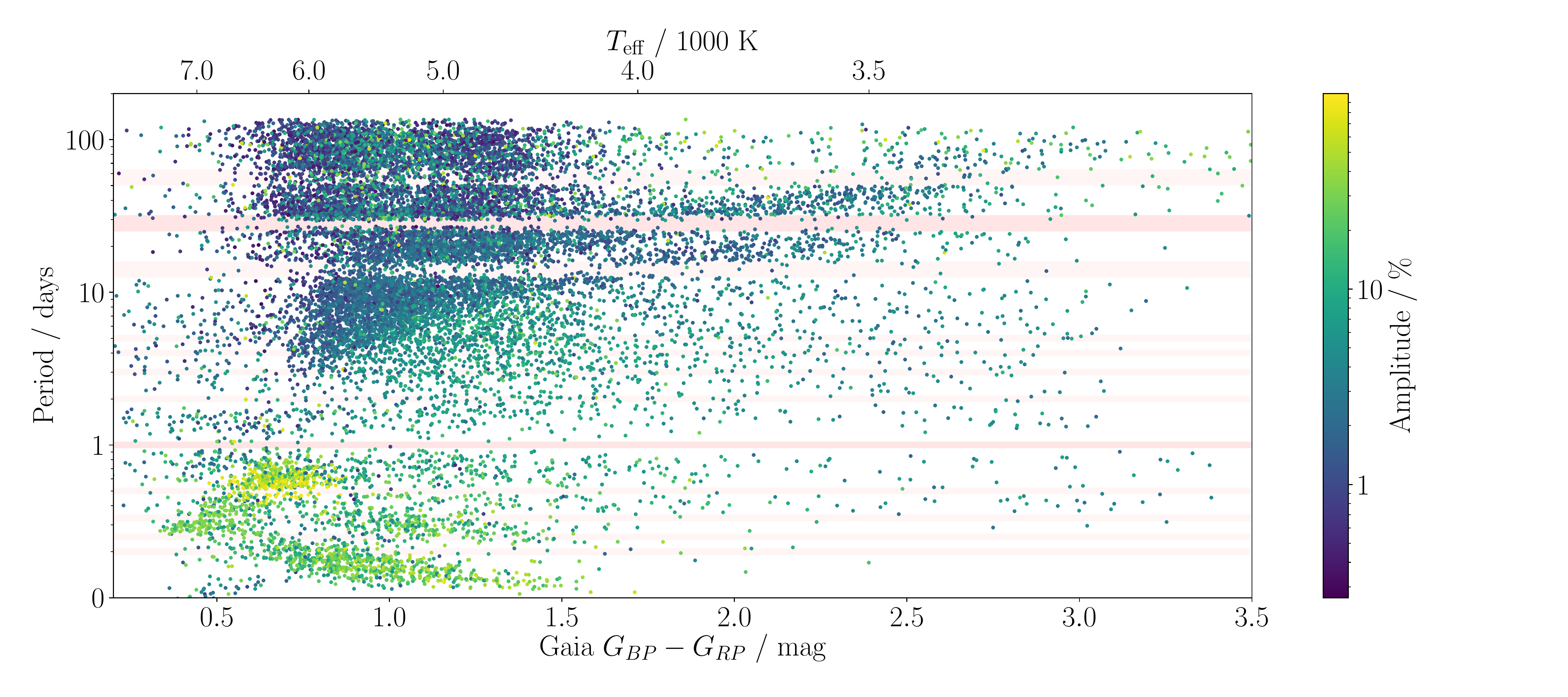}
    \caption{Effective temperature and \gaia{} \bprp{} colour against period for \nperiods{} stars. The colour indicates the $5^{\mathrm{th}} - 95^{\mathrm{th}}$ percentile spread of the signal in relative flux. To aid the eye, horizontal strips indicate regions of period space likely affected by systematics arising from the Moon or the 1-day sampling alias, with multiples of these periods more transparent.} 
    \label{fig:period_colour_mag}
\end{figure*}

{We plot our variability periods against colour in Figure \ref{fig:period_colour_mag} and see a number of prominent features.} Most striking is the high density of stars known in the literature as e.g. the `I-Sequence' \citep{Barnes2003} or the `Ridge' \citep{Kovacs2015} spanning a period range from $4 - 40$ days and \bprp{} $0.75 - 3.5$. The shape of this envelope has been empirically defined by \citet{Angus2019}, using a broken power-law gyrochronology model calibrated against the $\sim$\,800 Myr old Praesepe cluster.

We see a large number of long-period ($>40$ days) objects between \bprp{} of $\sim 0.7$ to $\sim 1.4$. {We would expect a higher density of detections at this colour range due to the high-density main-sequence turnoff and red clump, as shown in Figure \ref{fig:HR_diagrams}(b). Older main-sequence stars at this colour range may exhibit long period rotational modulation. The Cepheid instability strip lies within this colour range, and we would expect to see long-period oscillations from evolved stars driven by the $\kappa$ mechanism \citep{Saio1993}.}

Far below the I-sequence we see a high density of much shorter period, high amplitude variability amongst hot objects at \bprp{} $\sim 0.5 \rightarrow 1.5$, and Period $< 1$ day. This population corresponds to the top of the main sequence on an HR diagram.

We see two distinct groups of objects  in period range shorter than 1 day, trending to short periods  with increasing colour index (\bprp{} $~0.75 \rightarrow ~1.5$). The two distinct groups are from the same region of the HR diagram - the equal-mass binary main sequence. The light curves showed distinct eclipsing binary signals (as seen in object 1 in Figure \ref{fig:example_lcs}), however, the longer period branch contained light curves phase folded on the correct period and in the shorter period branch light curves phase folded on half this period. This is an artefact of the RMS minimisation step described in Section \ref{subsec:aliases}. For eclipsing binaries with slightly different primary and secondary eclipse depths the full period will show a `cleaner' phase folded light curve with separate primary and secondary eclipses. In comparison, for an equal depth binary
{the phase folded light curve will have a similar RMS if folded on the correct period or half the period, with the primary and secondary plotted over one another in phase space.}

Finally, we observe an increasing upper period envelope with increasing colour for \bprp{} $> 1.5$. We see a number of objects with \bprp{} $> 2.5$ having variability periods up to and exceeding 100 days. These objects are discussed in detail in Section \ref{subsec:mdwarfs}. 

\subsection{Period Bi-modality}
\label{subsec:results:bimodality}

Within the I-sequence envelope we see a hint of a region lacking in periodic signals between $\sim3500{}$K and $\sim{}4500$K (\bprp{}  2.5--1.5) and $\sim{}$15--30 days. This gap has been the topic of extensive discussion in recent papers (such as \citet{McQuillan2014, Davenport2018}), and although faint, we do observe this gap in this ground-based data set. This gap has previously been fitted using a gyrochrone, roughly following a $\text{\teff{}}^{1/2}$ relation \citep{Davenport2018}, as well as an empirical model using a similar $\text{\teff{}}^{1/2}$ relation \citep{Gordon2021}. 

\begin{figure}
    \centering
    \includegraphics[width=0.8\linewidth]{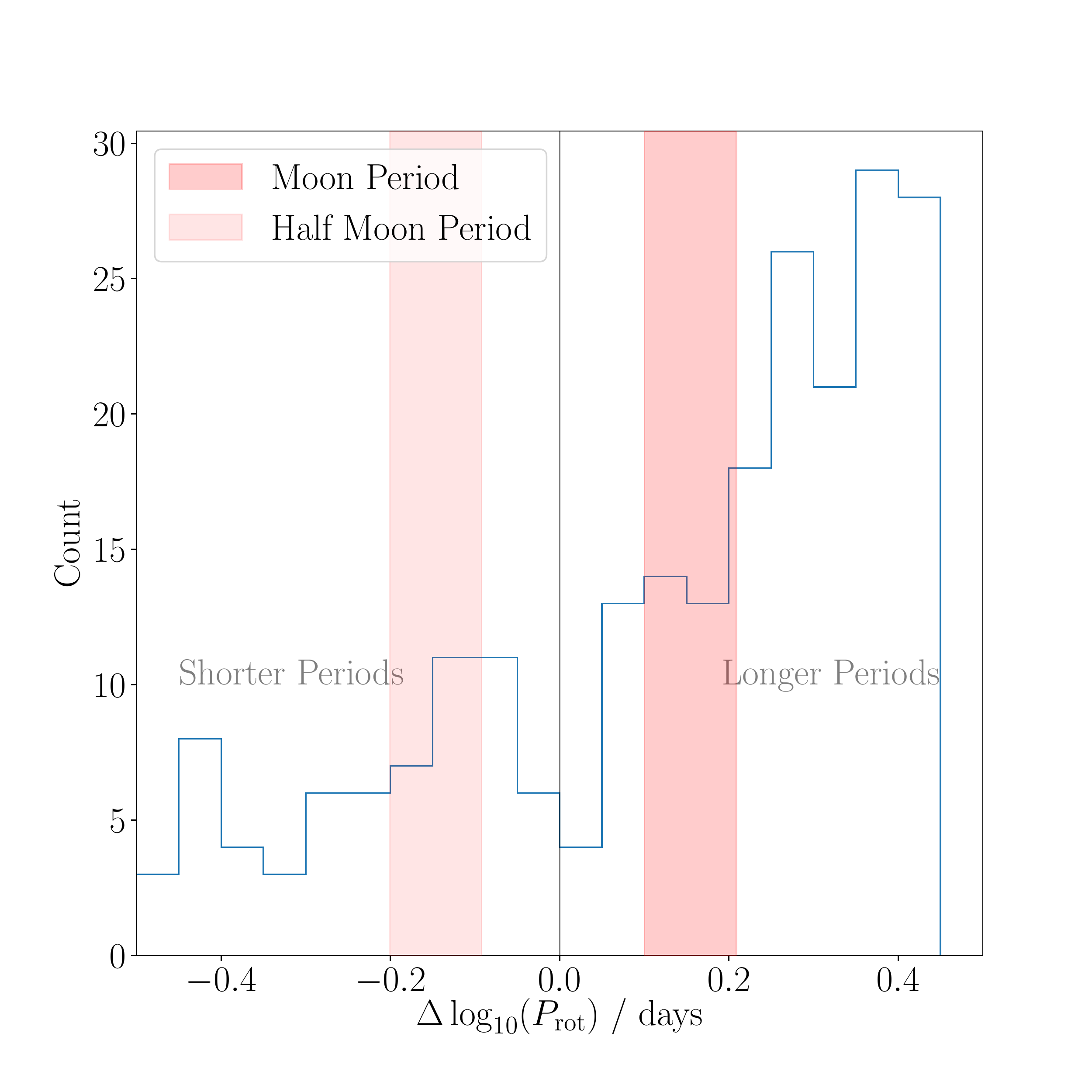}
    \caption{Distribution of the distance from a 600 Myr gyrochrone of the log periods for stars $1.4 <$ \bprp $< 2.2$. We see two peaks in the distribution, with a reduced number of rotation periods along the model gyrochrone {(grey vertical line)}. The range of distances from the model to the Moon and half Moon period are included to demonstrate the lower density of objects does not arise from a gap due to the Moon.}
    \label{fig:period_gap_histogram}
\end{figure}

To demonstrate the gap is present in our data we conduct the same analysis as in Figure 3 of \citet{Davenport2018}. We subtract model periods calculated with a 600 Myr gyrochrone defined in \citet{Meibom2011} from our periods. We constrain our data set to objects such that $1.4 <$ \bprp $< 2.2$ to avoid the gyrochrone crossing the Moon signal sampling gaps. In Figure \ref{fig:period_gap_histogram} we observe a dearth of objects along the gyrochrone, demonstrating the same gap as in the \kepler{} field is present within the \ngts{} data. 

\begin{figure*}
    \centering
    \begin{subfigure}[t]{0.45\textwidth}
         \centering
         \includegraphics[width=\textwidth]{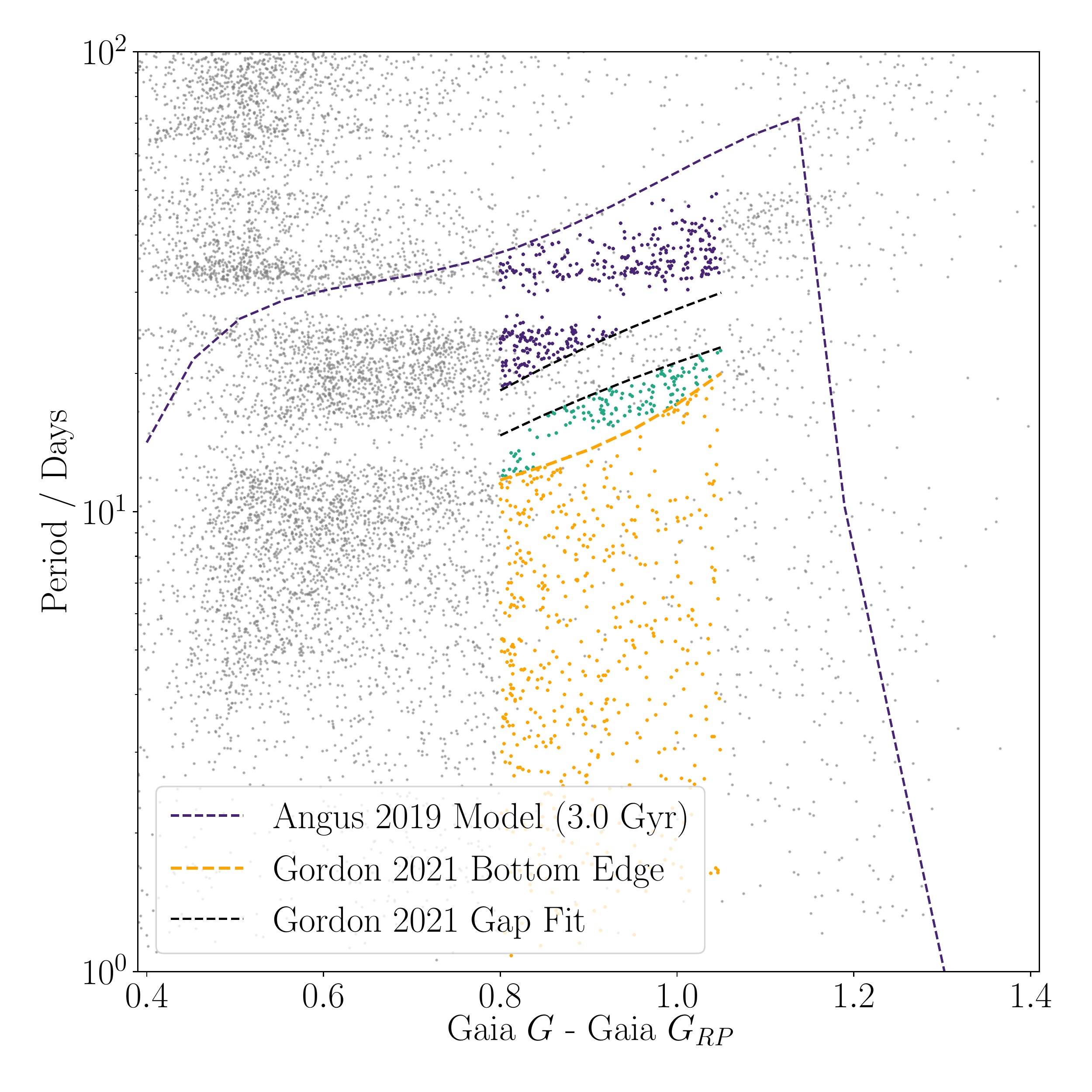}
        \caption{\gaia{} \grp{} colour vs. period. Three sub-samples spanning the observed period gap (above, below and significantly below the gap) are defined using models (see legend) and coloured blue, green and orange, respectively. Note we do not plot the large model uncertainties defined for the \citet{Angus2019} model for stars outside the range $0.56 < $ \bprp{} $ < 2.7$ ($0.31 < $ \grp{} $ < 1.17$).}
         \label{fig:3_samples_plot}
    \end{subfigure}
    \hfill
    \begin{subfigure}[t]{0.45\textwidth}
         \centering
         \includegraphics[width=\textwidth]{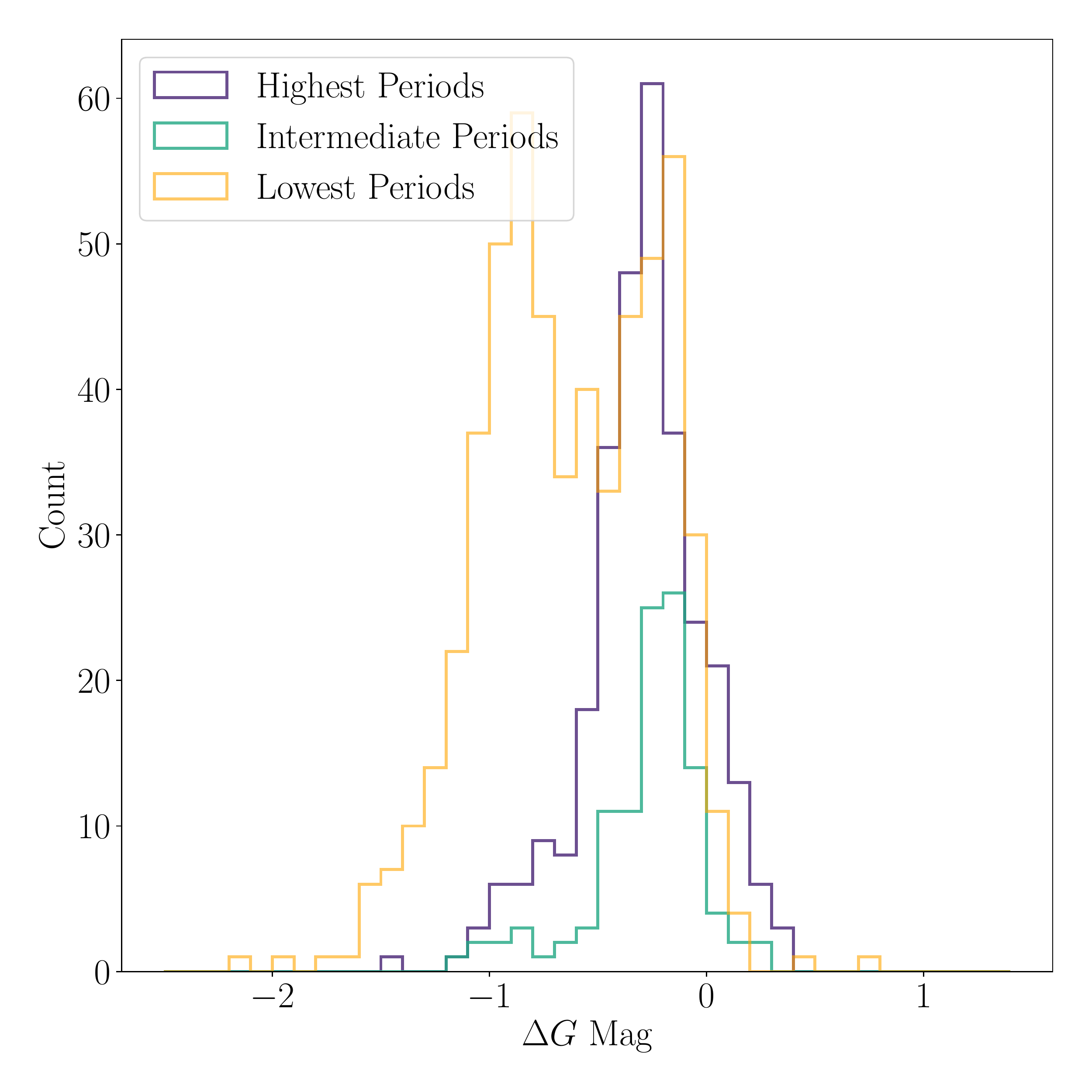}
        \caption{Histogram of distance in \gaia{} $G$ from a main-sequence isochrone (1 Gyr, Solar metallicity) for our three sub-samples (coloured as in panel (a)).}
         \label{fig:delta_g_hist}
    \end{subfigure}
    \caption{Panel (a): Period-colour diagram of our sample, with three sub-samples defined by empirical models from \citet{Gordon2021} and \citet{Angus2019}. Panel (b): Histograms of the magnitude difference in each of the three sub-samples from a main-sequence isochrone.}
    \label{fig:3_samples}
\end{figure*}

In Figure \ref{fig:3_samples} we separate our sample into three sub-samples based on a bi-modality gap model and empirical short-period lower limit from \citet{Gordon2021}. We observe how far these objects lie in absolute magnitude from an approximate main-sequence isochrone defined at 1 Gyr with Solar metallicity ($\Delta G$), as plotted in Figure \ref{fig:HR_diagrams}. We use this to assess where the three sub-samples lie {on the CMD}, to ascertain if they arise from distinct stellar populations in terms of colour and intrinsic brightness.
We elect to remove potentially evolved stars, giants and sub-giants to ensure the models from \citet{Gordon2021} and \citet{Angus2019} which are fitted to main-sequence stars from \kepler{} and \ktwo{} are applicable. We use the \textsc{evolstate} code described in \citet{Huber2017} and \citet{Berger2018}. The code gives crude evolutionary states for stars based on temperature and radius, with the models derived from Solar-type stars. We remove objects with the \texttt{`subgiant'} or \texttt{`RBG'} flags.

We define our 3 sub-samples using a number of model constraints in period-colour space. We use the fifth-order polynomial model defined in \citet{Angus2019} to constrain the long-period upper envelope of stars, and the edge-detection based fit from \citet{Gordon2021} to constrain the short-period lower envelope. 
We calculate the upper and lower edge of the gap using the model defined in \citet{Gordon2021}, and select stars from our I-sequence envelope on either side of this branch. This model was only defined for $0.8 < \text{\grp{}} < 1.05$, so we only use objects within this bound to define the sub-samples.
Our third sub-sample is defined as all objects below this boundary in period and will consist of stars not included in the \kepler{} and \ktwo{} data sets which fall well below the well defined I-sequence in period.
The model fits used in this section are detailed in Appendix \ref{app:model_params} and plotted in Figure \ref{fig:3_samples_plot}.

The histograms in $\Delta G$ plotted in Figure \ref{fig:delta_g_hist} show two similar single-peaked distributions from our two longer period sub-samples, and a distinct double-peak distribution for the shorter period sub-sample.
We note that this second peak lies approximately 0.75 magnitudes above the peaks of the two longer period sub-samples which could indicate a population of binary objects which is not present in the upper two sub-samples.
This confirms our previous observation from the HR diagram: a group of very short period objects just above the main sequence, which could correspond to a sample heavily contaminated by binary sources.
The two longer period sub-samples appear to have by-eye similar distributions of $\Delta G$, which leads us to believe the two branches are drawn from similar stellar populations in terms of colour, intrinsic brightness {and multiplicity}.

\section{Discussion}
\label{sec:discussion}

\subsection{Comparison to similar studies}
\label{subsec:comparison}

\begin{figure*}
    \includegraphics[width=\linewidth]{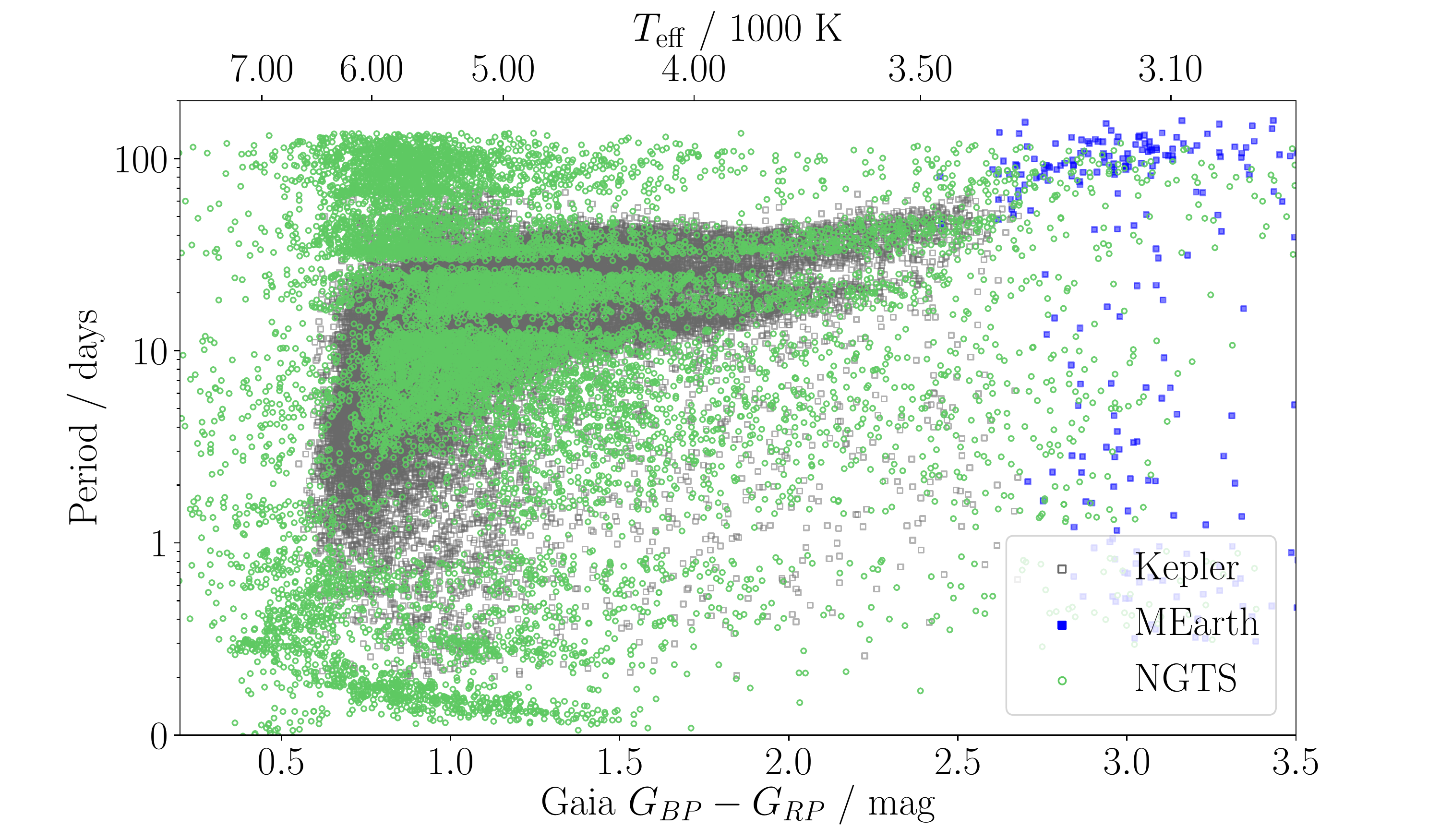}
    \caption{Effective Temperature vs Period data compared for this study (\ngts{} data, green circles), \citet{McQuillan2014} (\kepler{} data, grey squares) and \citet{Newton2018} (\mearth{} data, blue squares).}
    \label{fig:kepler_mearth_comparison}
\end{figure*}

The \ngts{} data set demonstrates that we are able to use ground-based photometry to conduct stellar variability studies matching the scale of space-based data. In contrast to, for example, the \kepler{} data set used by \citet{McQuillan2014} and \citet{Davenport2018}, \ngts{} sources are not pre-selected. This provides a much more representative sample of field stars which is demonstrated in the much higher number of objects which lie away from the high-density I-sequence envelope of stars in period-colour space. Objects which lie within the I-sequence will encompass a selection of stars most likely to be main-sequence, single objects similar to the \kepler{} input catalogue.
We overlay data from the \kepler{} rotation study by \citet{McQuillan2014} with our data in Figure \ref{fig:kepler_mearth_comparison}.
In particular, we see a high density of objects at \bprp{} $\sim 1.0$ with periods longer than roughly $40$ days not present in the Kepler data set. These objects lie in the RGB and AGB on the HR diagram, so will be giant objects which have not been removed from the \ngts{} study.
We also see a large number of objects with much shorter periods than the I-Sequence envelope. These objects lie above the main sequence {on the CMD} and will be either short-period binary sources or potential YSOs.

In addition to finding astrophysical signals of interest, we were also able to observe systematic periodicity down to amplitudes of $0.3\%$. 

This study highlights the power of ground-based photometric surveys in terms of the size and precision of the data set. We are able to extract a data set which rivals that of the \kepler{} and \ktwo{} missions, with a much longer baseline (in the case of \ktwo{}) and a much greater range of pointings (in the case of \kepler{}).
As a corollary, this study also serves as an exercise that ground-based photometric data may prove more difficult to analyse systematically than space-based data due to increased sources of noise and aliasing. We note a lower recovery rate of periodic signals than other studies. \citet{McQuillan2014} found variability in 25.6\% of their $\sim$\,$130,000$ objects, \citet{Gordon2021} found variability in almost 13\% of their $69,000$ objects, and \ngts{} was able to find variability in about 2\% of \nobjs{} objects. We note that 21\% of all objects were flagged as having signals arising from Moon contamination, our largest source of systematic noise in the study.

{The combination of a relatively long baseline ($\sim250$ days) and multiple pointings (94 used in this study) allows the NGTS data set to probe out to reasonably long period regimes (~0.1--130 days) and  across a range of spectral types (late-A to mid-M).}

\subsection{Long Period M-Dwarfs}
\label{subsec:mdwarfs}

Previous studies such as \citet{Newton2018} have used targeted ground-based photometry to extract very long period variability for M dwarfs. We also observe these extremely long periods ($>100$ days) in our M-dwarf sample. Figure \ref{fig:period_colour_mag} shows an upwards trend in period in the mid-M dwarf sample at $T_{\text{eff}}<3500K$. {In order to provide a useful comparison to the MEarth rotation study, we also assessed this trend for just dwarf stars (as defined by \texttt{evolstate})}. Our sample contains {751 non-evolved, dwarf} objects with variability periods with \gaia{} \bprp $>2.21$, which is the bluest limit of the MEarth rotation study {catalogue}.

In this study, the fields chosen had at most a 250-day time series, which allows us to robustly extract periods up to roughly 125 days in length. \citet{Newton2018} observed periods up to 140 days long for some of these objects, hypothesising that an upper limit close to this period would occur through Skumanich-like angular momentum loss for stars of the ages observed in the local thick disc. 
Using the Skumanich $t^{1/2}$ relation and taking the age of the local thick disc to be $8.7 \pm 0.1$ Gyr \citep{Kilic2017} we calculate the longest Skumanich relation period to be approximately $145$ days.
{The NGTS rotation periods qualitatively agree with the distribution of rotation periods seen in M dwarfs by \citet{Newton2018}, however, we reach the period limit of the NGTS data just shy of the $\sim 140$ day limit in the MEarth detections.}
It is interesting to note the Skumanich relation still appears to hold from the longest period objects across samples, even into the fully convective M-dwarf population for which the physics of spin-down is not fully understood.
Further observations of much older open clusters could shed light on this interesting long-period M-dwarf sample, and observations with much longer {time} baselines would allow us to probe into period regimes where spin-down could be more efficient than the Skumanich relation.
We note that current photometric space missions such as TESS \citep{Ricker2014} may be useful to shed light on this long term variability across the sky, but only at the ecliptic poles where objects will be observed for up to 1 year {continuously, with a one year gap before another year of continuous observation}. Most of the sky will only be observed for 28 days at a time, meaning a maximum of 14 day periods could be reliably extracted.

This \ngts{} study overlaps both the \kepler{} rotation period data and the \mearth{} rotation period data, allowing more robust comparisons to be made between the two previously disjoint samples. The \ngts{} data set provides a broad view into stellar rotation, targeting similar Solar-type stars as observed by \kepler{}, as well as more diverse populations across the HR diagram and across a range of pointings.

\subsection{Period Bi-modality}
\label{subsec:bimodality}

We continue the ongoing discussion regarding the rotation period gap \citep{McQuillan2014, Davenport2018, Reinhold2019, Reinhold2020, Angus2020, Gordon2021},
including the first ground-based data set to have observed this feature in period temperature space. Although the gap is not as clear as in the space-based data, we align models from a number of previous works to a region of lower density in the \ngts{} data, as shown in Figure \ref{fig:period_gap_histogram}. 

{By utilising empirical models from previous studies on \kepler{} and \ktwo{} data, we separated our sample into three sub-samples: this is seen in Figure \ref{fig:3_samples}.}
Within the two upper sub-samples, we see the highest period objects are on average further above the main sequence in $G$ than the lower period objects. This effect has been previously observed, as \citet{Davenport2018} saw a small increase in period as we move up in magnitude from the main sequence, but not as far as to be influenced by large numbers of binary objects. We note, similar to the \citet{Davenport2018} study that we do not account for metallicity or age when considering the distance from a Solar metallicity defined main-sequence isochrone at 1Gyr. Metallicity has been shown to affect the amplitude of variability signals and additionally may lead to observational biases whereby for a given mass, higher metallicity stars' variability is more easily detected \citep{See2021}. There is also the possibility of contamination by lower mass-ratio binary systems. 
Further observations of open clusters with defined stellar ages and a tight single-star main sequence may afford more conclusive evidence towards this period gradient across the main sequence. Such studies have been conducted on open clusters across a large range of ages such as Blanco 1 ($\sim 100$ Myr) \citep{Gillen2020}, Praesepe ($\sim 800$ Myr) \citep{Rebull2016, Rebull2017}, Ruprecht 147 ($\sim 3$ Gyr) \citep{Gruner2020} and M67 ($\sim 4$ Gyr) \citep{Barnes2016}.

The two sub-samples do not appear to be significantly contaminated by multiple systems and arise from similar locations on the HR diagram. Combined with the knowledge that these objects are from a range of pointings, this supports the conclusion of \citet{Gordon2021} that these two sub-samples do not derive from two distinct star formation epochs.

A broken spin-down law as discussed in \citet{Gordon2021} would be explained well by our data, including the possibility that the (very few) objects observed within this gap are currently transitioning between the two longer period sub-samples. In this broken spin-down law, the angular momentum change of the star will deviate from the expected $t^{1/2}$ relation proposed by \citet{Skumanich1972} due to the transfer of angular momentum between the envelope and the core. Prior to this transfer of angular momentum, the core and envelope are decoupled, resulting in the expected $t^{1/2}$ spin-down of the envelope but with a rapidly rotating core which will then reduce or even stop the spin-down once the core and envelope re-couple. 
This model has been suggested to fit \kepler{} data in addition to \ktwo{} data \citep{Angus2020, Gordon2021}, and theorists such as \citet{Lanzafame2015} and later \citet{Spada2020} have incorporated these effects into stellar evolution models which have been shown to fit observed cluster data of different ages. The proposed models include a two-zone model of internal stellar coupling, with a parameter describing the mass dependence of the coupling. 
The recent analysis of the $\sim 3$ Gyr old open cluster Ruprecht 147 by \citet{Gruner2020} demonstrates that the model from \citet{Spada2020} incorporating internal angular momentum transfer is best suited to model the rotational evolution of stars redder than K3 in comparison to more naive gyrochronology models. 

Another suggestion for the origin of this gap comes from an analysis by \citet{Reinhold2019} and \citet{Reinhold2020} of \ktwo{} data. In their proposed model, the gap arises from objects in which the photometric variability arising from spots and faculae is of similar magnitude, thus cancelling out {resulting in lower amplitude variability that is correspondingly harder to detect}. They observed a slight decrease in signal amplitude on either side of the gap in period, and hypothesised objects of this period could exhibit spot-faculae photometric cancellation. 
We do not observe such an obvious decrease in signal amplitude in our full sample, and when considering a smaller range of amplitudes more aligned with the \ktwo{} sample we again did not see this amplitude gradient. This may be attributed to NGTS photometry being less precise than Kepler, and a small change on a signal of $1\%$ amplitude may not be detectable.
To accurately determine the dominant surface feature of a star requires observations of spot-crossing events during planetary transits or Doppler images, neither of which are appropriate for follow-up from a large-scale photometric study.

\section{Conclusions}
\label{sec:conclusion}

In this study we extract robust variability periods for \nperiods{} stars out of \nobjs{} stars observed with the Next Generation Transit Survey (\ngts{}), based in Paranal, Chile. This is the largest ground-based systematic photometric variability study conducted to date with such precise and high-cadence photometry and highlights both the advantages of such studies as well as the challenges.
Using precise ground-based photometry, plus a generalisation of the autocorrelation function to irregularly sampled data, we are able to detect variability amplitudes down to levels of 0.3\%. 
The contamination of signals by systematics demonstrates that using ground-based photometry requires further thought than using much cleaner space-based data in order to avoid false positives arising through aliases. The most common source of aliases arose from Moon contaminated signals as well as aliasing from the 1-day periodic sampling intrinsic to ground-based observations. We demonstrate we are able to overcome these limitations and produce robust variability signals across our sample.

In comparison to previous large-scale stellar variability studies, we note that with \ngts{} we are able to observe across the Southern sky (in comparison to \kepler{}'s single pointing, as in \citet{McQuillan2014} and \citet{Davenport2018}). We do not pre-select our targets as is the case for \kepler{} and \ktwo{}, and so we are able to observe variability across a more varied stellar sample.
In particular, we extract long term variability periods for a population of cool dwarfs, similar to a population observed by \citet{Newton2018} using \mearth{}. This is made possible through our longer observation baseline than space-based missions such as \ktwo{}.
This large population, sampled across the sky over a long {(250 day) baseline} allows this study to connect previous space-based studies on main-sequence, predominantly Solar-type stars with ground-based M-dwarf studies, which were previously unconnected.

Within the bulk of our rotation period `I-Sequence', we observe a gap between 15 and 25 days, first observed by \citet{McQuillan2014}, and later studied in detail by \citet{Davenport2018}, \citet{Reinhold2019}, \citet{Reinhold2020}, \citet{Angus2020} and \citet{Gordon2021}. 
Using models from \citet{Gordon2021}, \citet{Angus2019} and \citet{Meibom2011} we are able to demonstrate that the gap is present in our data set, and also show that the two sub-samples of main-sequence objects above and below this gap appear to arise from similar stellar populations {on the CMD} which are not contaminated by high levels of binarity. This supports the hypothesis of a broken spin-down model as proposed by \citet{Lanzafame2015} and \citet{Spada2020} rather than distinct populations of star formation.

We also conclude that although a large population study of field stars is useful for assessing trends in the wider stellar population, without well-defined ages of target stars it is difficult to confirm angular momentum models. We suggest that studies of open clusters with well-defined ages and tight rotation sequences such as the recent study by \citet{Gruner2020} will yield the most conclusive evidence towards how stellar angular momentum evolves over the lifetime of a star.
Additionally, we observe a number of interesting non-main-sequence populations, including a small population of objects which lie well above the main sequence with short rotation periods. Follow-up observations of these targets would allow us to ascertain whether these stars are young, single stars such as T-Tauri objects, or multiple star systems.
This data set presents a wealth of additional data with many avenues for follow-up science. These include both continued systematic variability analysis of the \ngts{} data and also more in-depth analysis of interesting sub-populations of variable objects not explored in this cardinal \ngts{} variability study.

\section*{Acknowledgements}
{We thank the anonymous referee for constructive suggestions and comments that improved this paper.}
Based on data collected under the NGTS project at the ESO La Silla Paranal Observatory. The NGTS facility is operated by the consortium institutes with support from the UK Science and Technology Facilities Council (STFC) under projects ST/M001962/1 and ST/S002642/1. 

This work was performed using resources provided by the Cambridge Service for Data Driven Discovery (CSD3) operated by the University of Cambridge Research Computing Service (\url{www.csd3.cam.ac.uk}), provided by Dell EMC and Intel using Tier-2 funding from the Engineering and Physical Sciences Research Council (capital grant EP/P020259/1), and DiRAC funding from the Science and Technology Facilities Council (\url{www.dirac.ac.uk}).

This research has made use of the VizieR catalogue access tool, CDS, Strasbourg, France (DOI: 10.26093/cds/vizier). The original description of the VizieR service was published in A\&AS 143, 23

JTB is supported by an STFC studentship as part of the CDT in Data Intensive Sciences.
EG gratefully acknowledges support from the David and Claudia Harding Foundation in the form of a Winton Exoplanet Fellowship. MNG acknowledges support from MIT's Kavli Institute as a Juan Carlos Torres Fellow and from the European Space Agency (ESA) as an ESA Research Fellow.
JSJ greatfully acknowledges support by FONDECYT grant 1201371 and from the ANID BASAL projects ACE210002 and FB210003. JAGJ acknowledges support from grant HST-GO-15955.004-A from the Space Telescope Science Institute, which is operated by the Association of Universities for Research in Astronomy, Inc., under NASA contract NAS 5-26555. 

\section*{Data Availability}

The \ngts{} data underlying this article are publicly available through the ESO Archive in line with the \ngts{} data publication policy.

The variability period results produced in this article are available as supplementary material and at CDS via anonymous ftp to \url{cdsarc.u-strasbg.fr} (130.79.128.5) or via \url{https://cdsarc.unistra.fr/viz-bin/cat/J/MNRAS}.
% \textcolor{red}{[repository name, e.g. the Dryad Digital Repository], at https://dx.doi.org/[doi]}

%%%%%%%%%%%%%%%%%%%%%%%%%%%%%%%%%%%%%%%%%%%%%%%%%%



\bibliographystyle{mnras}
\bibliography{export-bibtex} % if your bibtex file is called example.bib

%%%%%%%%%%%%%%%%%%%%%%%%%%%%%%%%%%%%%%%%%%%%%%%%%%

%%%%%%%%%%%%%%%%% APPENDICES %%%%%%%%%%%%%%%%%%%%%

\appendix

\section{Detailed Moon Signal Analysis}
\label{app:moon}

\begin{figure}
    \centerline{
        \includegraphics[width=0.9\linewidth]{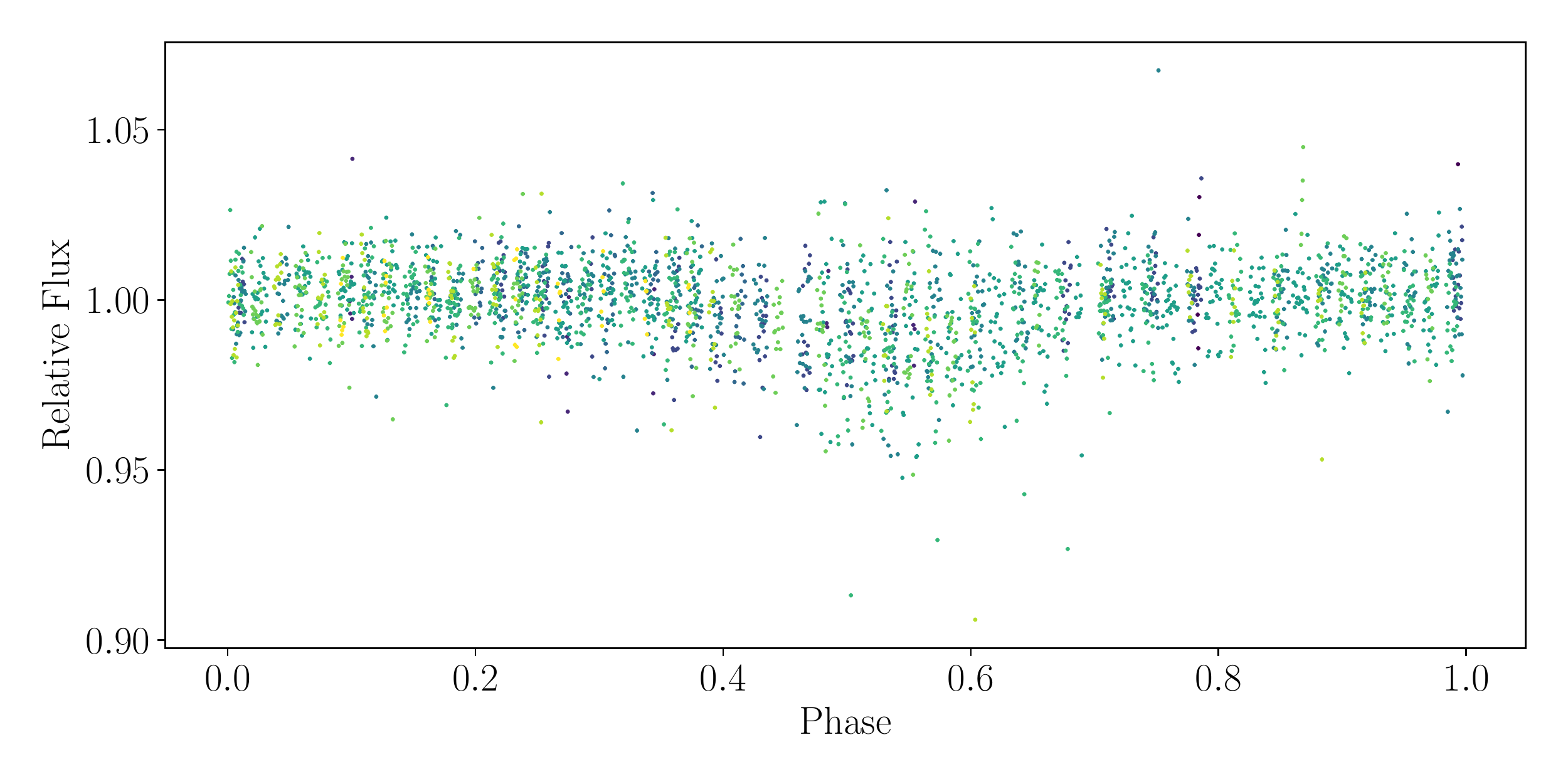}
    }
    
    \centerline{
        \includegraphics[width=0.9\linewidth]{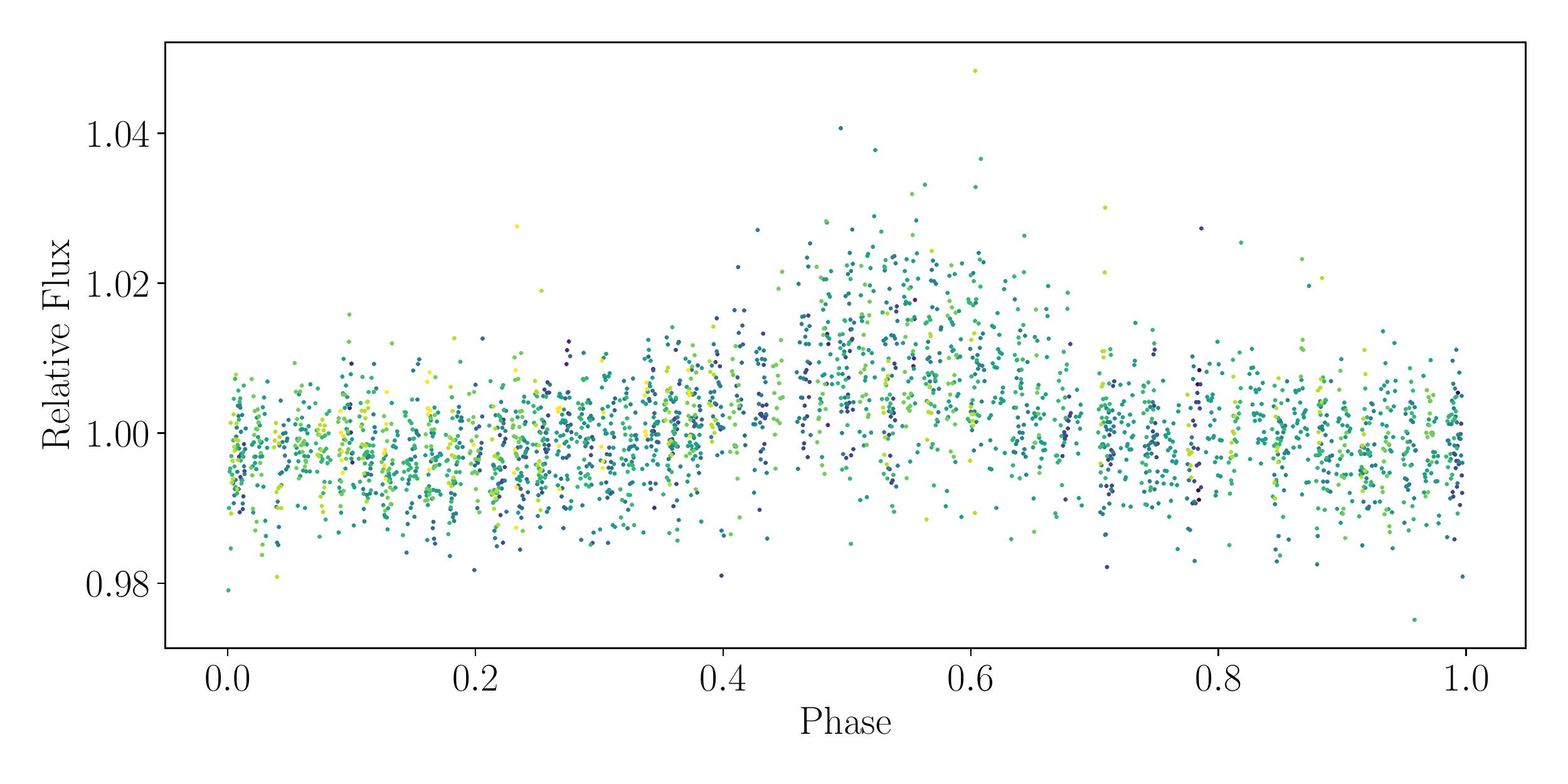}
    }
    \caption{Two examples of typical Moon tainted signals. For each object, the light curve is phase folded on the expected Moon period and epoch. 0.0 \& 1.0 phase are at new Moon, 0.5 phase is at full Moon. We see an example of an over-corrected signal with a typical decrease in flux at full Moon (top). An under-corrected signal demonstrates the opposite trend (bottom). Both signals exhibit an increase in scatter at full Moon, with an otherwise fairly flat light curve.
    \label{fig:moon_data}}
\end{figure}

In order to systematically detect Moon contaminated signals (for example as shown in Figure \ref{fig:moon_data}), we fit a model to the flux data, phase folded on the expected Moon period for each \ngts{}{} field. The expected Moon period is calculated from a scaled expected Moon brightness curve, calculated as a product of the on-sky separation of the field from the Moon and the Moon illumination fraction, $I = (1 + \cos{(\theta_{phase}})) / 2$. $\theta_{phase}$ is the Moon phase angle defined for a time and ephemeris. For most fields, this gave a period of approximately 28.5 days, between the synodic and sidereal periods as expected.

\begin{figure}
    \includegraphics[width=\linewidth]{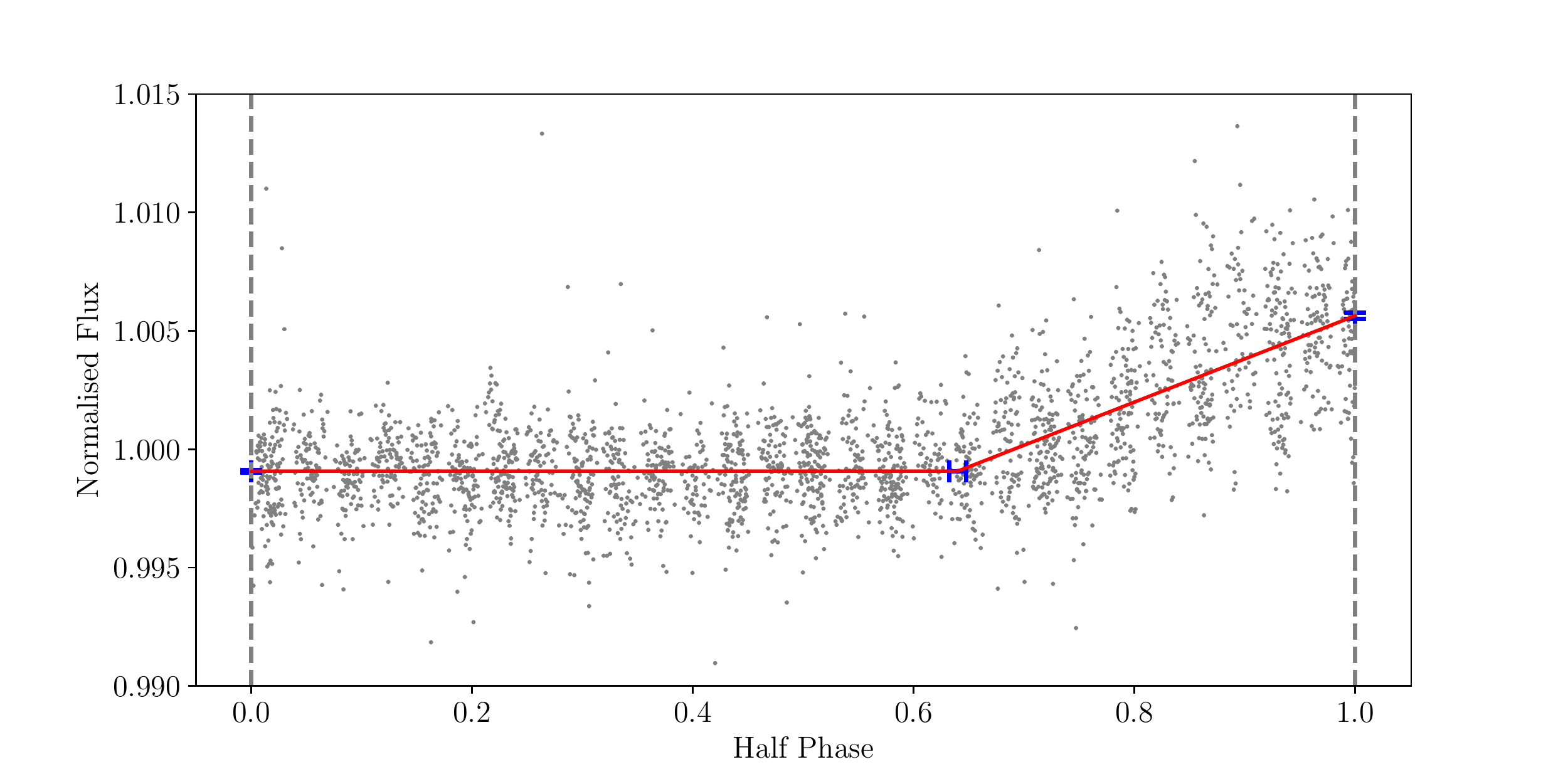}
    \caption{The three-parameter Moon model fit is used to assess if a signal is contaminated by the Moon. The flux data is phase folded on the period of the Moon and then again in half such that 0.0 in phase corresponds to new Moon and 1.0 in phase corresponds to full Moon.}
    \label{fig:moon_model_fit}
\end{figure}

The model is a simple three-parameter, piecewise model described in Equation \ref{eq:moon_model}, where the parameter $x$ is the location in half phase $x \in [0, 1]$.

\begin{align}
    \label{eq:moon_model}
    &\begin{cases} 
      \mathrm{flux}_0 & 0 \leq x \leq \mathrm{turnover} \\
      mx + c & \mathrm{turnover} < x \leq 1 \\
   \end{cases}
\end{align}
\noindent Where
\begin{align*}
   m &= \frac{\mathrm{flux}_1 - \mathrm{flux}_0}{1 - \mathrm{turnover}} \notag\\
   c &= \mathrm{flux}_1 - m \notag
\end{align*}

\noindent We fit for the 3 parameters $\mathrm{flux}_0$, $\mathrm{flux}_1$ and $\mathrm{turnover}$.
This model fit is assessed by checking the following criteria, with an example shown in Figure \ref{fig:moon_model_fit}.

\begin{itemize}
    \item Is the model turnover point at the expected point in phase? (between 0.2 and 0.8 in half-phase).
    \item Is there a flux RMS increase after the model turnover point?
    \item Is there a noticeable (i.e. $>1\sigma$) change in flux from new to full Moon?
    \item Is there any missing data at full Moon?
\end{itemize}

If 3 or more of these criteria are met, the object is flagged as Moon contaminated and removed from the processing.

\section{Model parameters}
\label{app:model_params}

In Figure \ref{fig:3_samples_plot} we use empirical models defined in \citet{Angus2019} and \citet{Gordon2021}. In this section, we provide the model equations and the parameters used.

\subsection{Angus model}
We use the Praesepe-calibrated gyrochronology relation defined in \citet{Angus2019}. The mathematical form of this fifth-order polynomial relationship is given in Equations \ref{eq:angus1} \& \ref{eq:angus2} below for two different \bprp{} regimes:

\begin{equation}
    \log_{10}(P_{\text{rot}}) = c_A\log_{10}(t) + \sum_{n=0}^4 c_n[\log_{10}(\text{\bprp{}})]^n
\label{eq:angus1}
\end{equation}

\noindent for stars with \bprp{} $< 2.7$ and

\begin{equation}
    \log_{10}(P_{\text{rot}}) = c_A\log_{10}(t) + \sum_{m=0}^1 b_m[\log_{10}(\text{\bprp{}})]^m
\label{eq:angus2}
\end{equation}

\noindent for stars with \bprp{} $> 2.7$. Here $P_{\text{rot}}$ is the rotation period in days, and $t$ is age in years. We use the best-fit coefficients from Table 1 of \citet{Angus2019}.

% \begin{table}
%     \centering
%     \caption{A table of model coefficients used in Equations \ref{eq:angus1} \& \ref{eq:angus2} as defined in \citet{Angus2019}.}
%     \begin{tabular}{@{}cc@{}}
%     \toprule
%     Coefficient & Value           \\ \midrule
%     $c_A$       & $0.65 \pm 0.05$ \\
%     $c_0$       & $-4.7 \pm 0.5$  \\
%     $c_1$       & $0.72 \pm 0.05$ \\
%     $c_2$       & $-4.9 \pm 0.2$  \\
%     $c_3$       & $29 \pm 2$      \\
%     $c_4$       & $-38 \pm 4$     \\
%     $b_0$       & $0.9 \pm 0.5$   \\
%     $b_1$       & $-13.6 \pm 0.1$ \\ \bottomrule
%     \end{tabular}
    
%     \label{table:angus}
% \end{table}

\subsection{Gordon Model}
We use the \ktwo{} calibrated model from \citet{Gordon2021} to define the upper and lower edges of the bi-modality gap seen in the I-sequence envelope. 
The gap edges are fitted using a function of the form:

\begin{equation}
    P = A(\text{\grp{}} - x_0) + B(\text{\grp{}} - x_0)^{1/2}
\label{eq:gordon}
\end{equation}

\noindent where $P$ is the rotation period in days. This equation is defined empirically for \ktwo{} stars with $0.8 <$ \grp{} $< 1.05$. We use the best fit coefficients defined in Table 7 of \citet{Gordon2021}.

The lower edge of the \ktwo{} sample from \citet{Gordon2021} used an edge-detection method, and as such no parametric model form was given. We instead define our lower edge by eye, taking the edge-detection fit line from the \citet{Gordon2021} paper.

% If you want to present additional material which would interrupt the flow of the main paper,
% it can be placed in an Appendix which appears after the list of references.

%%%%%%%%%%%%%%%%%%%%%%%%%%%%%%%%%%%%%%%%%%%%%%%%%%

% Don't change these lines
\bsp	% typesetting comment
\label{lastpage}
\end{document}